
\overfullrule = 0pt

\magnification=1200
%\font\cmc=cmcsc10

\def\gtorder{\mathrel{\raise.3ex\hbox{$>$}\mkern-14mu
    \lower0.6ex\hbox{$\sim$}}}
\def\ltorder{\mathrel{\raise.3ex\hbox{$<$}\mkern-14mu
    \lower0.6ex\hbox{$\sim$}}}
\def\pagenumbers{\footline={\hss\tenrm\folio\hss}}

\noindent\hangindent=40pt
\baselineskip=20 pt

% Some useful definitions and macros   "defs tex a"

%
%  Books..
%

%
%  Journals..........
%

\def\AsA#1#2{  {\it Astron. Astrophys.\/},       {#1}, #2.}

\def\AJ#1#2{   {\it Astron. J.\/},               {#1}, #2.}
\def\ApJ#1#2{  {\it Astrophys. J.\/},            {#1}, #2.}

\def\ApJS#1#2{ {\it Astrophys. J. Supp. Ser.\/}, {#1}, #2.}

\def\CJP#1#2{  {\it Canadian J. Phys.\/},        {#1}, #2.}

%
% ApJ's new definitions for some of these:
%
\def\AsA#1#2 {  A\&Ap, #1, #2.}
\def\AJ#1#2  {  AJ,    #1, #2.}
\def\ApJ#1#2 {  ApJ,   #1, #2.}

\def\ApJS#1#2{  ApJS,  #1, #2.}
\def\CJP#1#2 {  CJP,   #1, #2.}
%
%  Miscellaneous.....
%
%\def\etal{ {\it et al. }}

% #1 number of fig. e.g. '3b'   #2 text of caption
\def\ref{\noindent\hangindent=1.5cm\hangafter=1\relax}
% enter in {} e.g.:  Author, I. 1985.\MN{112}{246}
%
%  Extra Math Macros.
%

\def\ltapprox{\hbox{\raise.5ex\hbox{$<$}
\kern-1.1em\lower.5ex\hbox{$\sim$}}}
\def\gtapprox{\hbox{\raise.5ex\hbox{$>$}
\kern-1.1em\lower.5ex\hbox{$\sim$}}}

\def\overmag#1.#2 {\hbox{$#1\spm$\kern -6.0pt .\kern 3.0pt #2}}
\def\overday#1.#2 {\hbox{$#1\spd$\kern -4.0pt .\kern 2.0pt #2}}

% ...................
%--------------------
\nopagenumbers
\baselineskip=12pt
 
\centerline{\bf DYNAMICS OF BROAD EMISSION-LINE REGION IN NGC~5548:}
\centerline{\bf HYDROMAGNETIC WIND MODEL versus OBSERVATIONS}
\bigskip

\baselineskip=12pt
\centerline{Mark Bottorff, Kirk T. Korista \& Isaac Shlosman}
\centerline{\it Department of Physics and Astronomy}
\centerline{\it University of Kentucky}
\centerline{\it Lexington, KY 40506-0055}
\centerline{\it E-mail: bottorff@pa.uky.edu, korista@pa.uky.edu, shlosman@pa.uky.edu}
\smallskip
\centerline {and}
\smallskip
\centerline{Roger D. Blandford}
\centerline{\it Theoretical Astrophysics 130-33}
\centerline{\it California Institute of Technology}
\centerline{\it Pasadena, CA 91125}
\centerline{\it E-mail: rdb@tapir.caltech.edu}
\bigskip
\smallskip
\vfill

\baselineskip=12pt
\centerline{\bf ABSTRACT}
\smallskip

We analyze the results of long-term observations of broad-line region (BLR) 
in the Seyfert~1 galaxy NGC~5548 and provide a critical comparison with the 
predictions of a hydromagnetically-driven outflow model of Emmering, 
Blandford and Shlosman. We use this model to generate a time series 
of C~IV line profiles that have responded to a time varying continuum. 
Our modifications to the model include cloud emission anisotropy, cloud 
obscuration, a CLOUDY-generated emissivity function and a narrow-line component 
which is added to the BLR component to generate the total line profiles. The 
model is driven with continuum input based on the monitoring campaigns of 
NGC~5548 reported in Clavel et al. and Korista et al., and the line 
strengths, profiles and lags are compared with the observations. 

Our model is able to reproduce the basic features of C~IV line variability in this
active galactic nucleus, i.e., time evolution of the profile shape and 
strength of the C~IV emission line {\it without varying the model 
parameters}. The best fit model provides the effective size, the dominant geometry, 
the emissivity distribution and the 3D velocity field of the C~IV BLR and constrains 
the mass of the central black hole to $\sim 3\times10^7\ {\rm M_\odot}$. 
The inner part of the wind in NGC~5548 appears to be responsible for the
anisotropically emitted C~IV line, while its outer part remains dusty and
molecular, thus having similar spectral characteristics to a molecular torus, 
although its dynamics is fundamentally different.

In addition, our model predicts a differential response across the C~IV line
profile, producing a red-side-first response in the relative velocity interval of 
$3,000\ {\rm km\ s^{-1}}$ to $6,000\ {\rm km\ s^{-1}}$ followed by the
blue mid-wing and finally by the line core. Based on the comparison of data and 
model cross-correlation functions and one and two-dimensional transfer functions, 
we find that the rotating outflow model is compatible with observations of the BLR 
in NGC~5548.

\bigskip\noindent
{\it Subject headings:\/} black hole physics --- galaxies: active ---
galaxies: individual (NGC~5548) --- galaxies: nuclei --- galaxies: Seyfert 
--- line: formation
 
\vfill\eject

\baselineskip=20pt

\pagenumbers
\pageno=1
\line{\bf 1. INTRODUCTION\hfill}
\bigskip

Observations of active galactic nuclei (AGNs: quasars and Seyfert galaxies) 
have revealed broad emission lines superimposed on a nonstellar continuum 
(Osterbrock 1996 and refs. therein). 
Understanding broad emission-line regions (BLRs) of AGNs is clearly among the outstanding
problems of modern research on galaxy evolution. Since their discovery, extensive 
effort has been invested in understanding
the BLR geometry, its relationship to the central engine in AGN and to the host
galaxy. In the heart of the problem lies the origin of gas in the BLR and its
physical state and dynamics. The emerging picture is that of photoionized
clouds moving within $\sim 1$ pc of a central UV-to-X-ray continuum
source. The observational support to this view consists of a rough proportionality 
between the emission lines and continuum, the inferred diverse ionization 
states of the emitting gas, and the similarity between the equivalent 
widths of hydrogen lines from the BLRs. Further constraints on the
physical conditions within the emitting clouds and their kinematics come from the 
observed relative line ratios, line profiles and their variability. 

Although the resolution of the overall problem of BLRs has proven to be elusive 
due to a large phase space of possible solutions ill-constrained by
observations, a consensus has emerged on a number of issues. Firstly, the
BLR clouds should be numerous as their line profiles appear to be
smooth; the individual clouds are smaller than the central continuum
source because deep intrinsic Lyman edges have never been observed. The
clouds are optically thick at the Lyman limit, in Ly$\alpha$ and possibly other 
strong resonance lines, such as C~IV $\lambda$1549 (e.g., Netzer 1990).  Secondly, 
systematic differences between the high and
low-ionization emission line  (HILs and LILs) profiles, e.g., full-width at 
half-maxima (FWHM) and velocity shifts between line peaks
(Gaskell 1982; Wilkes 1984; Collin-Souffrin and Lasota 1988; Espey et al. 1989;
Corbin 1990 and others), have suggested
ionization stratification in the BLRs. Thirdly, the results of recent
monitoring campaigns of a number of AGNs have suggested much smaller sizes of
line-emitting regions than previously anticipated and independently
confirmed the ionization stratification of the BLRs (see review
by Peterson 1993). Both the continuum and the emission lines in
AGNs are observed to vary with time. The emission line
variability lags behind and represents the BLR response to the
underlying continuum variations. The response or lag time
differs from line to line. Blandford \& McKee (1982) proposed a
``reverberation mapping'' technique which uses the line and continuum 
variability data to infer the geometry and kinematics of the BLR. A number
of phenomenological models utilizing simple geometries and motions of the
ionized gas within the BLR have been calculated (e.g., Welsh and Horne
1991; Perez, Robinson and de la Fuente 1992a,b; Robinson 1995).
The long-term continuous monitoring 
and high quality data produced in recent observations supply additional 
constraints on the theoretical models of the BLRs which can be tested now 
(Gondhalekar, Horne and Peterson 1994 and refs. therein).  
 
All published models of BLRs fall within one or more of the following kinematic categories: 
infall, outflow, rotation and random motion. It is understood
theoretically, that if the AGNs are accretion-powered, a strong inflow must be
present, although it is by no means clear whether this
has anything to do with the observed motion of the ionized gas in the BLRs. One 
possibility is that the inflowing gas stays molecular for as long as it is capable 
of shielding its dust from the external radiation field and internally-generated heat.
The dynamics of ionized gas, which is subject to a
partial or full radiation field from the central continuum source, may be affected
by the radiation force as well. Various acceleration mechanisms for 
the ionized gas in the BLRs have been proposed, such as continuum radiation
pressure-driven clouds (Matthews 1986 and refs. therein), thermally-driven 
winds (Weymann et al. 1982; Cassidy and Raine 1996), line-driven winds from 
accretion disks (Shlosman,  Vitello and Shaviv 1985; Murray et al. 1996), and
magnetically-driven winds (Emmering, Blandford and Shlosman 1992; K\"onigl
and Kartje 1994). Alternatively, radial 
infall models have been tested in order to reproduce the observed emission line 
profiles (Krolik and London 1983), sometimes accompanied with random cloud 
motions (Kwan and Carrol 1982). Rotating accretion disk models of BLRs have been 
explored as well (e.g., Shields 1978; Matthews 1982). A recent revival of
the line-driven winds from accretion disks as being responsible for the formation of
the broad lines in AGNs (Murray et al. 1996; Chiang and Murray 1996) has revealed 
certain advantages of this model, as well as some outstanding 
observational and theoretical difficulties (e.g., de Kool and 
Begelman 1995). 

Models of BLRs involving clouds on purely radial (inward/outward) or Keplerian orbits
appear to be inconsistent with recent observations of variability
in the broad emission line profiles of Seyfert galaxies (e.g., Peterson et al. 1993). 
It is also difficult to 
understand the absence of angular momentum in the (radially) inflowing clouds.  
Further difficulties accompany the cloud model itself. Among these
are the stability of radiatively accelerating clouds against internal radiation 
pressure gradients and the accompanying hydrodynamical shear, the cloud confinement 
mechanism, and the source of BLR gas (e.g., Matthews and Capriotti 1985). 
Exceedingly large masses for the central black holes, $\gtorder 10^9\ {\rm M_\odot}$,
are required to account for the observed line wings, especially in the brighter
AGNs, if these lines are broadened by cloud motion alone. 
On the other hand, both cloudy and continuous gaseous disk models for the BLRs 
possess their own share of unusual requirements. Among these are, irradiation of 
the disk from high altitude, a large radial range to account for the
entire line profile, disk warping, inflow rates and low gas temperatures which may 
lead to Jeans instabilities in the outer disk and possibly to star formation. Also, 
disk clouds are expected to be short-lived, being sheared on the orbital timescale. 

Emmering, Blandford and Shlosman (1992, EBS) proposed a new interpretation
of BLRs. In particular, it was suggested that BLRs in AGNs are associated with
hydromagnetic (MHD) winds from dusty molecular accretion disks. In this picture,
dense molecular
clouds are loaded on the magnetic lines threading the disk, and are centrifugally
accelerated outward and above the disk by magnetic stress, like beads on a 
rotating wire. When exposed to
the central UV continuum source, the clouds are quickly photoionized and produce
emission lines, until the clouds dissipate. The EBS model introduced a 
self-consistent velocity field above the disk, which combines both rotation and
outflow. The accretion disk itself was considered to be opaque and to extend 
beyond the wind radius. Within this framework, both HILs and LILs originate
in the wind, at different lattitudes and radii. The EBS model accounts also for 
a transverse confinement of emitting clouds by an ambient magnetic field 
({\it cf.} Rees 1987), thus removing a number of long-standing 
problems associated with presence of hot and dense intercloud medium in pressure 
equilibrium with the clouds (e.g., Fabian et al.\ 1986; Mathews \& Ferland 1987). Furthermore, EBS found that in order to model the broad line 
profiles, the BLR clouds need to originate only over two decades in disk radius, 
if electron scattering by a hot, $\sim 10^6$ K, gas contributes to the line 
broadening in the wings. The importance of hydromagnetic winds for AGN unification
schemes was further emphasized by K\"onigl and Kartje (1994), while de Kool and
Begelman (1995) studied implications of magnetized winds for broad absorption-line
quasars. 

In this paper, we propose to test the MHD model in
the best studied case of a variable AGN --- the Seyfert~1 galaxy NGC~5548.  
We model the dynamics and geometry of the BLR in this object, 
constraining it through the observed variable C~IV emission line,
its correlation with changes in the central ionizing continuum and the
response across the C~IV line profile. Based on the modeled response to the 
observed continuum variations, our best fit model provides the effective size, 
the dominant geometry, the emissivity distribution in the velocity space, and 
makes certain predictions about the response of the C~IV emission line in NGC~5548. 
The underlying dynamical model of the BLR is that of a hydromagnetic 
wind from an accretion disk. By incorporating a number of
additional assumptions outlined below and in section~3, we show that the 
3D velocity field in the EBS model having both outflow and rotational components 
can explain the observed profiles and their variability in this object. 

A number of specific issues guided us when applying the EBS model
to the BLR in NGC~5548. Firstly, based on the line response to the variable
ionizing continuum, the cloud population that is responsible for the C~IV
emission must be optically thick with hydrogen column densities
in excess of $10^{23}\ {\rm cm^{-2}}$. This follows from the observed
C~IV emission which has responded positively to the continuum change. No
negative response, i.e., decrease in the C~IV flux with increase of the central
continuum flux, has ever been detected. The prevalence of optically thick clouds 
in NGC~5548 is also supported by the He~II line
emission which originates close to the central source and varies
with a substantial amplitude and in near tandem with the central continuum. A direct
consequence of large column densities in the BLR clouds is that their C~IV
emission should be strongly anisotropic (Ferland et al. 1992).
Secondly, the overall geometry of the BLR in the EBS model is disk-like. Hence
a number of optically thick clouds are expected to
overlap in a particular direction, causing cloud-cloud obscuration of
the central UV continuum source. Continuum optical depth effects must,
therefore, be included. Lastly, the C~IV emission line appears to be the dominant 
and the most uncomtaminated line 
in the broad line spectra of a large number of AGNs. For this reason we limit
our discussion to this line only. Although our analysis is applied directly to the 
BLR in NGC~5548, it is possible that it has much wider consequences for the
unified model of AGNs.

In section 2, we provide a brief summary of the monitoring campaigns of 
NGC~5548. Section 3 gives a general overview of the EBS model, including the 
modifications that were introduced and the values of the parameters that were 
used in the present modeling 
of the C~IV line variation in NGC~5548. In section 4, the results of 
the model calculations are presented and section 5 compares the model with 
the data and makes a number of predictions. Additional comments and conclusions 
are given in section 6.

\bigskip
\line{\bf 2. OBSERVATIONAL SUMMARY ON NGC~5548\hfill}
\bigskip

NGC~5548 is a bright (V$\sim$13.5) and variable Seyfert~1 galaxy.
Because of these characteristics it has received considerable attention
in recent multi-wavelength ground and space-based campaigns to monitor
its continuum and emission line variability in the hopes of uncovering
the nature of both. Early investigations into the emission line
variability of NGC~5548 include Peterson (1987), Peterson, Korista \&
Cota (1987), Stirpe, de Bruyn \& van Groningen (1988), Maoz et al. (1990),
Peterson et al. (1990), Rosenblatt \& Malkan (1990), Wamsteker et al. (1990), and
Koratkar \& Gaskell (1991). While useful results were obtained, these
data were generally less than optimal in temporal resolution and
sampling in determining the physical parameters governing the broad
emission line gas, such as geometry, kinematics and emission line
responsivity. Thus ambitious monitoring campaigns were launched. In
1988-1989 the International Ultraviolet Explorer ({\it IUE}) observed
NGC~5548 every 4 days for 8 months (Clavel et al.\ 1991, hereafter
C91); accompanying the ultraviolet data were ground-based optical data
spanning the same time period (Peterson et al. 1991). Reliable lags
between the emission lines and the continuum were determined for the
first time, and crude velocity integrated transfer functions were
derived for several emission lines (Krolik et al. 1991; Horne, Welsh \& Peterson 1991; Dietrich et al. 1993; Maoz et al. 1993). A still more ambitious monitoring program which
combined data from the Hubble Space Telescope ({\it HST}), the {\it IUE}, and
from ground-based observatories was conducted in the Spring of 1993 to
address questions requiring higher signal-to-noise, spectral and
temporal resolution spectra. In this case NGC~5548 was observed once
per day in the ultraviolet for 39 days with the $HST$ and once every
other day for 75 days with $IUE$, again with accompanying ground-based
optical spectra (Korista et al. 1995, hereafter K95). The high quality
$HST$ data will be used to constrain our model, while data from both of
these campaigns will be considered in our analysis, below. 
 
The primary results of interest here from these two campaigns were: (1)
the ionization parameter and density of ionized gas vary across the
BLR and, hence, the BLR is stratified in ionization, at least in this object, 
with the highest ionization lines (N~V and He~II) forming mainly near to 
the source of ionizing photons ($\sim2$ light-days) and lines of decreasing
ionization forming increasingly further away; (2) the velocity field of
line emitting gas is not purely radial. No unambigous differential 
{\it lags} have been measured
across the C~IV $\lambda$1549 profile. With low signal-to-noise (S/N)
and marginal temporal resolution, the {\it IUE} spectra of the 1988-89
campaign do not lend themselves to convincing quantitative profile
analyses, while insufficient temporal coverage hamper the analyses of
the high temporal resolution, high S/N emission line profiles from the
1993 $HST$ campaign (K95). However, the observed variations in both
data sets do suggest that the C~IV profile responds differentially. For
example, K95 concluded that there were indications that the wings lead
the response of the cores of C IV, which would imply virial cloud motions.
Evidence for the presence of any radial motions have heretofore favored
inward flows.  Crenshaw and Blackwell (1990) attributed the {\it
redward asymmetric response} of the C~IV emission line profile during a
dramatic decline in the continuum in 1989 to the presence of inflow.
Done \& Krolik (1996) also attributed to a modest gravitational inflow
component a similar response in the red wing of C~IV to a dramatic
increase in the continuum during the 1993 $HST$ campaign. However, the
interpretation of these two isolated cases of redward asymmetric
response in C~IV (it was not seen in other similar continuum
variations), is not clear.  Wanders et al.\ (1995) chose not to
interpret it, in the context of the classical inversion of the transfer
function equation, and suggested a non-stationary origin of the red
wing response. Wanders \& Peterson (1996), in their study of the
long-term variations in the H$\beta$ emission line profile, concluded
that the large-scale profile variations in that line are not driven by
short time-scale reverberation effects. Unfortunately, the nature of
broad emission line profile variability remains unclear.

In spite of this, some statements have been made in the recent
literature regarding the nature of the flow of broad emission line
gas.  Based on analyses of the K95 spectroscopic data, Done \& Krolik
claimed to have excluded several kinematical models, among them the
rotating hydromagnetic wind model of EBS, although they soften their
claim in scenarios for which the far side of the BLR is obscured by,
for example, the putative accretion disk.  In view of a prevailing trend
in the literature to interpret the observations in terms of inflow, it
is our intention here to demonstrate that the outflow picture, based on
the EBS model of hydromagnetic wind, makes predictions which are
entirely consistent with the observations of the BLR in NGC~5548.
 
\bigskip
\line{\bf 3. GENERAL ASPECTS OF THE MODEL\hfill}
\bigskip

\line{\bf 3.1. Hydromagnetic Flow in the Broad Line Region\hfill}
\medskip

Using asymptotic solutions of hydromagnetic flows from accretion disks and the 
line-broadening effect of electron scattering, EBS were able to reproduce 
typical static UV line profiles for BLRs in AGNs. In this section, the EBS model is 
further elaborated in order to account for the temporal effects detected in the 
broad C~IV line of NGC~5548. We give only the necessary details of the
model and discuss modifications we have introduced. More can be found in EBS.

The EBS model assumes that an outflow originates from an optically thick Keplerian 
accretion disk of a dusty molecular gas. Thus an observer will see only the nearer 
hemisphere of the BLR. Magnetic field lines thread the disk and are frozen into the
ionized gas component there (Blandford and Payne 1982). 
Gas is occasionally loaded on the magnetic lines and centrifugally accelerated above 
and away from the disk. The clouds in this flow are 
assumed to be exposed to the central UV continuum source and to contribute to 
the line emission between the polar angles $\psi_l$ and $\psi_u$ above the plane of 
the disk. The innermost radius of the hydromagnetic wind in the disk plane, 
$r_{0,min}$, is defined by the transition from molecular to ionized disk which 
exists interior to this point. This radius is established by the sublimation 
temperature of the dust, $T_{max}\sim1,800$ K (Phinney 1989), and is given by 
EBS 
$$ r_{0,min} = 6\times 10^{16} L_{44}^{1/2} \zeta^{1/2} \phi^{-1/2} T_{m18}^{-2}\
     {\rm cm},     \eqno(1) $$
where $T_{m18}\equiv (T_{max}/1800\ {\rm K})$; $L = 10^{44}L_{44}\ {\rm ergs\ 
s^{-1}}$; $\zeta (<< 1$ at low latitudes above the disk) is the attenuation factor
of UV radiation; and $\phi\sim 0.1-1$ is the mean IR extinction efficiency (e.g.,
Sanders et al. 1989; Shlosman and Begelman 1989). Because the characteristic 
timescale for variability of the central UV-soft-x-ray continuum luminosity 
$L(t)$ ($\sim$ tens of days) is much shorter than 
accretion timescale ($> 10^3$ yrs), $r_{0,min}$ is fixed by the (unknown) historical 
maximum of the 
central luminosity and is not expected to respond to shorter timescale 
continuum variability. In other words, the only way to replenish the dust at some
radius in the disk destroyed by a `flash' in $L(t)$, is by moving the dusty gas 
inwards from a larger radius. This can be done only on the accretion timescale 
--- the longest timescale in the problem.
Because the historical maximum of $L(t)$ is not known and
in compliance with the observed He~II response, we fix $r_{0,min}$ at one light 
day, taking $(L_{44}\zeta/\phi)^{1/2} = 0.045$ in eq.~(1).

The EBS model provides scaling relationships for the density in the MHD
flow, the magnetic pressure, density, mass and size of individual clouds, with the 
position in the wind. For example, at the base of
the wind, the density is $\propto r^{-b}$, the magnetic energy density
$\propto r^{-(b+1)}$, and the mass loss per decade in $r$ is $\propto
r^{-(2b-3)/2}$, where $b$ is a constant. Dependence of physical parameters
on the position along the streamline is more complicated and explained
in EBS and Blandford and Payne (1982). 

\bigskip
\line{\bf 3.2. Emission Model of the Broad Line Region\hfill}
\medskip

Two important modifications have been introduced to the emission line model used
by EBS. Firstly, because clouds are optically thick to the Lyman continuum and 
sufficiently thick to C~IV line photons (Netzer 1990; Ferland et al. 1992), we 
introduce a simple emission anisotropy, $A(\omega)$, for this line
$$\hskip 1. in  A(\omega) =
  \cases{
  {\rm cos}\ \omega\hskip 1. in {\rm for}\ \  0\le\omega<\pi/2\cr
   \cr
  0\hskip 1.235 in {\rm for}\ \ \pi/2\le\omega\le\pi\cr}, \hskip 1.2 in (2)$$
where $\omega$ is an angle between the direction to central source and
the direction to the observer, as viewed from the cloud. Eq.~(2) means that
we ignore emission from the backside of the BLR clouds, which heavily
biases the observed line radiation to come from the far side of the BLR (Fig.~1).
This simple form accentuates the influence of anisotropy on line formation.
Although Ferland et al. calculated front/back C~IV intensity ratio from a
cloud being $\sim3:1$, possible dust opacity was not taken into account. For
the large column density clouds used here, destruction of C~IV photons by
the dust (if present) in the neutral part of the cloud will eliminate much of the C~IV
flux from the backside of the cloud.

Secondly, we use a simple approximation to the continuum optical depth effects in
the wind. This is done by assuming that all clouds are optically thick
to the ionizing continuum and that foreground clouds shield the
background clouds.  For the sake of simplicity, we neglect the
continuum radiative transfer through the clouds, taking them to be of
an infinite optical depth at all energies relevant to C~IV.  This is identical 
to having a pure geometrical obscuration. The cloud obscuration is defined as
$$ \tau_{obs} = \int\sigma_c n_c dR,     \eqno(3) $$
where the integral is taken along a spherical radius-vector $\vec R$, starting at 
the innermost wind boundary $r_{min}$ at a height $z$ above the disk plane; 
$\sigma_c$ and $n_c$ 
are the cloud cross-section and volume number density, respectively. Note, that
in our notation, capital $R$ is used for the spherical geometry, while small $r$
denotes a cylindrical radius. Eq.~(3) 
can be evaluated analytically using variables of the self-similar MHD wind model:
$$ \tau_{obs} = 3 \gamma F(\psi) R_{0,min} [(R/R_{0,min})^{1/3}-1], \eqno(4) $$
where $R_{0,min}$ is the spherical radial coordinate of the innermost obscuring 
clouds on the ray $\vec R$; $\psi$ is defined as an angle between $\vec R$ and the 
disk surface; $F(\psi)$ is an auxiliary dimensionless function (Fig.~2) which 
provides the polar dependence of the MHD model. 
$\gamma = 0.29$ lt-d$^{-1}$ is a normalization constant, that sets the 
strength of the cloud-cloud obscuration effect in the BLR. The choice of $\gamma$ 
determines the size of the emitting BLR and comes about from fitting the model C~IV 
line profile to the data (see section 4). The value of $\gamma$ chosen in this paper 
bounds the C~IV emitting region to be within 24 light days of the continuum source. 
Clouds lying at $\tau_{obs} \geq 1$ are completely obscured from the ionizing UV
continuum and, therefore, do not 
emit C~IV line radiation. We do expect that hard X-rays will penetrate well beyond
this boundary and hence the $\tau_{obs} \geq 1$ region might be responsible for LILs 
formation. Cloud obscuration can further enhance the ionization stratification 
in the BLR.

The C~IV volume emissivity of the wind in the direction of the observer is then given 
approximately by
$$ \hskip 1. in \eta = 
   \cases{
   \epsilon n_c\sigma_cA(\omega)(1- \tau_{obs})\hskip 1.3 in {\rm for}\ \ 
   \tau_{obs}\le 1\cr
   \cr
   0\hskip 2.455 in {\rm for}\ \ \tau_{obs} > 1\cr}, \hskip .42 in (5) $$
where $\epsilon$ is the surface C~IV emissivity of a cloud. The quantity $\sigma_c(1-\tau_{obs})$ then represents the area of a cloud that is exposed 
to continuum radiation and responds by emitting $\epsilon$. 

Based on the cloud parameters provided by the EBS model, one can, in principle,
calculate a self-consistent C~IV volume emissivity distribution in space.
However, such a calculation is impractical, because $\epsilon$ is a function of the 
incident continuum flux, the cloud column density and the gas particle density
in the cloud. Constraining the model parameters by using the C~IV line profiles
then would involve iterations in the multi-dimensional parameter phase space.
Instead, the cloud surface emissivity at a given ionizing flux, $\Phi(H)$, is
taken to be a weighted mean of surface emissivities of optically thick clouds
distributed uniformly in $\log n(H)$, as explained below. Here $\Phi(H)$ 
is the photon number flux above 1 Ryd which is incident on the cloud and $n(H)$ 
is the hydrogen number density of a cloud. The calculations were performed
using the photoionization code CLOUDY (Ferland 1994) and the radiative transfer 
was terminated when the electron fraction fell below 0.1. In this 
picture, the cloud
parameters are not finely tuned, rather a mixture of clouds with a wide range
of physical characteristics is assumed to exist at each $R$, subject to selection
effects described in the ``locally optimally-emitting cloud'' model (Baldwin et al.
1995).

The cloud hydrogen densities used in CLOUDY calculations ranged over six orders 
of magnitude, i.e., 
$10^8-10^{14}\ {\rm cm^{-3}}$, and their column densities ranged over $10^{20}-10^{24}\ 
{\rm cm^{-2}}$. This amalgam of cloud properties is consistent with those prescribed
by the EBS model.
%(Note that the range in cloud parameters applicable to our model 
%is much narrower,  $10^{8.5}-10^{12}\ {\rm cm^{-3}}$ and $10^{22}-10^{24}\ {\rm 
%cm^{-2}}$, see section 3.4.)  
An analytical fit to this surface emissivity is
$$ {\rm log}~\epsilon (C~IV) = 1.044[{\rm log}~\Phi_{18}(H)]^{0.67} + 8.0 \hskip .5 in 
     {\rm for}\ \ 1.0 < {\rm log}~\Phi_{18}(H) < 5.0, \eqno(6)$$
where $\Phi_{18}(H) = \Phi(H)/10^{18}\ {\rm cm^{-2}\ s^{-1}}$.
The shape of the continuum used for the CLOUDY simulations was that of the best 
fit found by Walter et al. (1994) to {\it IUE} and {\it ROSAT} data of NGC~5548. 
We scaled the 
variations in luminosity to the variations of continuum window observations of 
the {\it IUE} and {\it HST} reported in C91 and K95. The luminosity
was normalized by the estimated bolometric luminosity of $2.69\times 10^{44}\ {\rm erg\ 
s^{-1}}$ assigned to JD 2,449,116, using Hubble constant $H_0=75\ {\rm
km\ s^{-1}\ Mpc^{-1}}$ and deceleration parameter $q_0=0.5$. 
This results in a photon number flux of
$$ \Phi_{18}(H) = {2.139\times 10^{3}\over R_{16}^2}f_\lambda(t)\  {\rm cm^{-2}\ 
        s^{-1}},        \eqno(7) $$
where $R_{16}=(R/10^{16}\ {\rm cm}$), and $f_{\lambda}(t)$ is the normalized 
continuum window energy flux at a Julian Day $t$ in ${\rm erg\ cm^{-2}\ s^{-1}}$.
We normalized  $f_{\lambda}(t)$ by the 1460\AA\ continuum window energy flux of
$3.30\times 10^{-14}\ {\rm
erg\ cm^{-2}\ s^{-1}}$ on JD 2,449,116 for the K95 campaign, and by the 1350\AA\ window
energy flux of $3.62\times 10^{-14}\ {\rm erg\ cm^{-2}\ s^{-1}}$ for the C91 campaign. 
The line flux contribution at each
frequency can be calculated following the prescription given in section
4 of EBS, correcting for anisotropy of line emission and for finite continuum
optical depth effects. Because the BLR clouds move nonrelativistically,
we limit the accuracy of our calculations to the first order in $v/c$.
 
To explain the broad emission line profiles, the EBS model invoked both Doppler 
broadening, arising from the motion of individual clouds in the wind, and
electron scattering by a hot gas. No assumption about the pressure equilibrium
between the cold clouds and hot electrons was made whatsoever. Within this 
framework, the line cores are formed by unscattered line photons, while the line
wings result from electron scattering by a gas at $\sim 10^6$ K. Two models for hot
gas distribution have been considered by EBS, i.e., an accretion disk corona 
(so-called non-local scattering model) and an {\it in situ} gas surrounding emission 
clouds (local model). We use the latter model (described in section 5.4 of EBS) 
for our calculations as it is simpler and its physics is more transparent.  
EBS have shown that the gas at this range of temperatures is thermally 
stable in the presence of a confining magnetic field.

\bigskip
\line{\bf 3.3. Building the Model\hfill}
\medskip

A list of model parameters is provided in Table~1a with the adopted values. In 
addition
to the parameters listed in Table~1a are the following which are used
to generate model C~IV emission line profiles: (1) the UV-continuum
light curve which drives the model C~IV response, (2) a constant narrow
C~IV emission line component.

The model was constrained by the high-quality C~IV emission line
profiles from the {\it HST} campaign. To generate these time variable model
emission line profiles, the $\lambda$1460\AA\ continuum window light curve
from the {\it HST}-portion of the K95 campaign was used. To cover the time period
preceeding the {\it HST} observations, we used a simple functional fit to the {\it 
IUE} continuum. These data were handled 
by piecewise linear interpolation between data points.  The input continuum is 
shown in Fig.~3a. We also generated synthetic data
using the 1350\AA\ continuum window of the 1988-1989 {\it IUE} campaign
(C91). In this case, piecewise linear interpolation was also
used (Fig.~3b).
   
C~IV and other high ionization lines of NGC~5548 consist of a broad
(FWHM$\sim 5,000\ {\rm km\ s^{-1}}$), time variable component and a
time stationary component with FWHM$\sim 1,000\ {\rm km\ s^{-1}}$ that
rides on top of it. The region that produces the varying broad
component is the BLR and the stationary part is formed in the narrow
line region (NLR). We add a stationary NLR component to the synthetic
line profiles produced by the model. For the best fit, we chose a
Gaussian profile with a FWHM$ = 1,300\ {\rm km\ s^{-1}}$,
$F(C~IV)=5.9\times 10^{-13}\ {\rm erg\ cm^{-2}\ s^{-1}}$, and its peak
located at a velocity offset of $180\ {\rm km\ s^{-1}}$ blueward of the
rest frame of the central continuum source taken at redshift of 0.0174. 
The adopted NLR parameters were
close to those found from independent measurements of the K95 and
Crenshaw et al. (1993) {\it HST} data, namely the $F_{NLR}(C~IV)=6.62\times 
10^{-13}\ {\rm erg\ cm^{-2}\ s^{-1}}$, the FWHM$ = 1,100 \pm 100\ {\rm
km\ s^{-1}}$, and a peak offset of less than $200\ {\rm
km\ s^{-1}}$.

\bigskip
\line{\bf 3.4. Constraining the Model\hfill}
\medskip

Since the EBS model employs a self-similar solution to
the MHD flow, there is no {\it a priori} scale imposed on quantities
such as the gas density or the column density of a cloud. In order to
model the BLR in NGC~5548, it was necessary to establish an absolute scaling.
One option was to normalize the emission model by choosing $n_c$ and $\sigma_c$ 
in eq.~(5) at the innermost cylindrical radius of the BLR, $r_{0,min}$. However,
we have shown in section~3.1 that the model depends on their product 
rather than on each of these quantities separately. This means that
the fundamental parameter of the model is the characteristic differential 
obscuration $\gamma$ introduced in eq.~(4) as the cloud-cloud obscuration per 
unit length. Ideally, specifying $\gamma$ and the distribution
of cloud surface emissivities fully constrains the emission model,
allowing one to calculate the total C~IV line profile from the BLR and to compare it
with the observed profile. Unfortunately, uncertainties in the Hubble constant
and the distance to NGC~5548, anisotropy in the ionizing continuum, assumptions
used in calculating $\epsilon$, etc., all make such a comparison almost useless. 
Instead, in order to overcome these difficulties, we normalize the modeled C~IV 
profile to the observed one by requiring that the flux between relative
velocities $+3,000\ {\rm km\ s^{-1}}$ and $0\ {\rm km\ s^{-1}}$ of the
model be equal to the flux over the same relative
velocity interval in the data. The reason for this choice is that this
velocity interval is free from absorption or contaminating emission. We normalized the 
model line flux on JD 2,449,121 which was five days after the normalization 
of the continuum, chosen to match the $4.6$ day lag between
the C~IV emission line and the 1350\AA\ continuum window to the nearest day (K95).  
The adjustable parameters of the model were then tuned to give 
a fit to the rest of the data in the line profile segments of the {\it HST} data 
(defined in section~4.2.2).

The first step in constraining the model was an inspection of model line 
profiles superimposed on the observed line profiles of C~IV in K95. The model 
line shapes are most affected by the velocity field and the
emissivity of the BLR. We have used the basic parameters of the MHD
model in EBS in order to constrain the velocity field in the C~IV
emitting BLR of NGC~5548. The parameters of the EBS models have been explored 
in searching for the best fit to the C~IV line profiles. Adjustments to
the EBS model parameters reflect our search for proper boundary
conditions for the outflow as well as the uncertainty related to the
aspect angle of the external observer. The adopted parameters (Table~1a)
are close to the values given in the original EBS paper. 
The parameters of the MHD model
used here are as follows: the initial angle of magnetic field lines with the disk, 
$\theta_0$; the dimensionless total specific angular momentum, $\lambda$, in the
MHD flow --- a conserved quantity (eq.~[3.13] of EBS); the MHD flow 
density power law index $b$ and the cloud mass power law index $\alpha$. 
The geometrical parameters are the aspect 
angle $i$ of the observer with the rotation axis of the accretion disk; 
and the upper and lower boundary polar angles of the BLR, $\psi_l$ and $\psi_u$, 
as measured 
from the continuum source.  The electron scattering assumes a temperature of $T_e$ 
in the intercloud gas and the fraction of photons that scatter at least once in
the intercloud gas is given by $f_s$. 

Fixing the above parameters of the model allows one to calculate the C~IV line
profiles leaving only one degree of freedom in choosing their FWHM. Fitting
the observed C~IV line profiles then constrains the mass of the central
black hole, $M_{BH}$, to be $\sim3\times 10^7\ {\rm M_\odot}$. We also note, 
the physical meaning of $b=1.5$ in the best fit model is that the mass-loss rate 
per decade of disk radius $r$ is independent of $r$. 
 
The following features must be accounted for to allow a
meaningful comparison between the synthetic and observed C~IV lines:
(1) several absorption components in the C~IV profile, and (2) a small
correction in the flux of the extreme red wing of C~IV due to
contamination by other emission. Corrections for both absorption and
contaminating emission to the measurements of the data were based on the 
differences between the direct
integrations and fit to the C~IV emission line (described in section~4.2.2;
see also Fig.~9 of K95). 

\bigskip
\line{\bf 4. MODEL RESULTS\hfill}
\bigskip

This section is devoted to a direct comparison between the available data
on C~IV emission line in NGC~5548 and our model for the MHD wind from an
accretion disk. Using the observed UV continuum light curve as the driver, we 
compute the response of the synthetic C~IV emission line profile {\it without
varying the model parameters}. We 
compare the observed variable line profiles of K95 with the synthetic 
line profiles of the model. Finally, 
based on the cross-correlations between different segments of the line, we 
estimate the response times for different parts of the synthetic C~IV emission 
line profile and compare to the observed lags. 

\bigskip
\line{\bf 4.1. Model C~IV Line Profiles\hfill}
\medskip

Figure 4 provides sample line profiles generated by the model
superimposed on the {\it HST} data profiles. The
vertical axis shows the intensity in arbitrary units and the horizontal
axis is scaled in units of relative velocity offset from the center of
the C~IV line. The profiles chosen for this figure are at 5-day
intervals, starting with JD 2,449,097. A number of distinct absorption
features is clearly evident on the blue side of the observed lines
which have {\it not} been corrected.  The line profiles 
are quite similar in shape. The model
initially slightly overpredicts the red flux, but the data and model
curves merge together within ten days. In both campaigns, we find that the synthetic 
line profiles have a constant blue asymmetry of $\sim 8$\% in the wings (using the EBS definition), at 10\% peak line intensity. However, the asymmetry is time-variable in 
the core. At half-maximum the asymmetry changes from being 5\% red to 2\% blue, 
during the C91 campaign. During the K95 period, it changes from 0\% to 6\% in the red. 

\vfill\eject
\bigskip
\line{\bf 4.2. C~IV Line Light Curves\hfill}
\medskip

In this section we compare the temporal variations in the C~IV line flux between 
the model and the data.

\bigskip
\line{\bf 4.2.1. Integrated Line Fluxes\hfill}
\medskip
 
We first compare the integrated line fluxes for C~IV. In Fig.~5, the
light curve for the integrated line flux for the model is superimposed
on the data for {\it HST} portion of the K95 campaign.  The model curve was
created by driving the EBS model with a continuum input scaled by the
variations of the 1460\AA\ continuum window as described in section~3.2.
In both the data and the model, we have integrated over a velocity
interval of $\pm 9,000\ {\rm km\ s^{-1}}$.  This interval was chosen
in order to avoid contamination by He~II and
other emission lines on the red side of the line.

The model follows the general upward trend of the data. The amplitudes
are similar and both the model and the data show the same concavity at
the beginning and at the end of the time interval, so the large scale
variations are reproduced by the model. 

Fig.~6 shows a similar plot for the C91 campaign.  This time, the model
curve was created by driving the EBS model with a continuum input
scaled by the variations of the 1350\AA\ continuum window during the
C91 campaign.  The integrated lines in this figure represent the total
flux in the relative velocity interval $\pm 10,840\ {\rm km\ s^{-1}}$.
The plots in the figure show that the model reproduces three rise and
fall events of the data and most of the time the model is within 10\%
of the data.  This is remarkable because the model was  normalized to
the absolute value of the flux in the red core on JD 2,449,121 during
the K95 campaign and yet provides a reasonable match to the C91 campaign
as well, which happened four years earlier!

\bigskip
\line{\bf 4.2.2. C~IV Line Segments\hfill}
\medskip

Looking only at the integrated line flux hides variations that may be
occuring in different parts of the line.  It is, therefore, advantageous
to dissect the line profile into a number of segments and study each
separately.  In K95, the C~IV line profile was divided into 4 distinct
segments, i.e., red and blue cores and red and blue wings. We find it more 
insightful to divide the line into 6
segments instead. The line segments are defined in the Table~1b in terms of Doppler
velocity (i.e., relative velocity) as follows, the FBW or far blue wing 
($-9,000\ {\rm km\ s^{-1}}$ to $-6,000\ {\rm km\ s^{-1}}$), the MBW or 
mid blue wing ($-6,000\ {\rm km\ s^{-1}}$ to $-3,000\ {\rm km\ s^{-1}}$), 
the BC or blue core ($-3,000\ {\rm km\ s^{-1}}$ to $0.0\ {\rm km\ s^{-1}}$), 
the RC or red core ($0.0\ {\rm km\ s^{-1}}$ to $3,000\ {\rm km\ s^{-1}}$),
the MRW or mid red wing ($3,000\ {\rm km\ s^{-1}}$ to
$6,000\ {\rm km\ s^{-1}}$) and  the FRW or far red wing ($6,000\ {\rm
km\ s^{-1}}$ to $9,000\ {\rm km\ s^{-1}}$). Note that in this model most of the line
emission in the far-wings and some of the emission in the mid-wings consist 
of photons which have been electron-scattered from the line cores and 
the mid-wings, respectively (see sections~3 and 5).

The flux variability in each of the above segments is shown in
Fig.~5, where the model curves are superimposed on the data.  A number of 
absorption features can be clearly identified in the
observed C~IV emission line profile (K95): two weak (Galactic)
interstellar absorption systems (C~IV $\lambda$1549\AA\ and Si~II 
$\lambda$1527\AA) in the blue
wing, and at least one semi-broad intrinsic system due to C~IV
blueshifted $\sim700\ {\rm km\ s^{-1}}$ relative to the line peak. The data
curves have been corrected for absorption and emission contamination in an
approximate way.
The corrections were based on a comparison between the direct integration and 
fits of the line segment light curves presented in K95.
The RC and the MRW had no corrections while the BC has been corrected for 
intrinsic absorption in the same way as in K95. 
Finally, 2/3 of the K95 correction for the Galactic absorption in the blue wing was 
applied to the MBW and the remaining 1/3 was applied to the FBW.

The FRW was handled differently than the other line segments because
it is significantly contaminated with the He~II emission which, in comparison to C~IV,
varies rapidly with time. Fe~II is another possible contaminant.
It is not possible at this time to estimate what
fraction of the original K95 correction between $3,000-10,840\ {\rm km\ s^{-1}}$ 
should be applied to our FRW, i.e., between $6,000-9,000\ {\rm km\ s^{-1}}$, or
to estimate its accuracy. Therefore, we simply adjusted the light curve in the data 
FRW until its mean coincided with that of the model FRW. This amounted to a shift
of $-15$\%. None of the subsequent analysis was based on the FRW.

In the cores, the data RC and BC have virtually the same amplitude
variations, and likewise with the model RC and BC (Fig.~5). In the mid and
far-wings, there is a blue asymmetry in both the data and the model line profiles, 
slightly larger in the latter. 

We have made a quantitative comparison of the model line segments with
the data line segments by computing the RMS fractional deviation for
each line segment during the {\it HST} portion of the K95 campaign.  The RMS
fractional deviation, $f$, is defined by 
$$    f^2 = {1\over N}\sum_i (X_i - M_i)^2/X_i^2, \eqno(8)$$
where $X_i$ and $M_i$ are the integrated line fluxes for the data and
the model during the $i$th observation, and $N$ is the total number of
observations. Thus, $f$ is a measure of the magnitude of a typical
fractional difference expected between the model and the
data at any time.  The computed values are as follows, $f_{FRW}=0.09$,
$f_{MRW}=0.08$, $f_{RC}=0.03$, $f_{BC}=0.03$, $f_{MBW}=0.08$ and
$f_{FBW}=0.07$. All model line segments
are within 10\% of the data.  The fractional deviation of the integrated 
line for the entire K95 is $f_{tot}=0.02$. We note, that while the model
reproduces successfully the general trends of the observed line variations
(Fig.~6), the synthetic light curves do not follow the short-time scale,
small-amplitude variations. The model limitations are further discussed in 
section~5.5.

%\vfill\eject
\bigskip
\line{\bf 4.3. Cross-Correlation Analysis of the C~IV line\hfill}
\medskip

With a good temporal resolution it is possible to investigate
correlations between an emission line (or portions of an emission line)
and the continuum.  A measure of the correlation is determined by the
analysis of the cross-correlation function, hereafter CCF (e.g., Krolik
et al. 1991).  This function is defined as
$$ C_{lc}(\tau) = {1\over{\Delta t\sigma_l\sigma_{cont}}}\int [F_l(t)-{\bar F_l}]
      [F_c(t-\tau)-{\bar F_c}]dt, \eqno(9) $$
where $F_c(t)$, ${\bar F_c}$ and $\sigma_{cont}$ are the continuum light
curve, the temporal mean of the continuum light curve, and the standard
deviation of the continuum light curve, respectively. $F_l(t)$, ${\bar
F_l}$ and $\sigma_l$ are the line emission light curve, the temporal
mean of the line emission light curve, and the standard deviation of the
emission line light curve.  $\Delta t$ is interval of
time during which the continuum and the line are observed.  The value
of $C_{lc}$ is the correlation coefficient of the line light curve with
the continuum light curve temporally shifted by a time interval
$\tau$.  For two curves that are well correlated, the correlation
coefficient will be close to one and the CCF will have a sharp, well
defined peak. The value of $\tau$, at which the correlation is maximal,
is called the response time or lag time of the line with respect to the
continuum.  Another measure of the response time is the CCF {\it
centroid} at some fraction of the maximum height of the CCF. The
centroid is useful when the peaks of the CCF are rounded, making the
identification of the peak difficult. In addition, the centroid of the
CCF is mathematically identical to the centroid of another function
called the one-dimensional transfer function (1DTF). The latter indicates how
the line strength changes when the system is driven by a continuum which
is a delta function in time.  We address the 1DTF later in section 5.
In this paper we present both the peak lag and the centroid lag for all
reported response times.  In practice, the definition of
$C_{lc}(\tau)$ was not used because monitoring campaigns are discrete in 
time.  A discrete analog of $C_{lc}(\tau)$, as developed by White and 
Peterson (1994), is used instead.

The lag time provides an important measure of the size of the BLR by 
interpreting the lag time as
being due to the light travel from a central continuum source to the
line-emitting region. The differing lag times for different emission
lines have provided the most direct evidence for the BLR being
stratified, with the highest ionization lines coming from the innermost
regions of the BLR and the lowest coming from the outermost regions.
Comparing the lag times of individual segments of a single
line with the continuum enables us to study the kinematic structure of
the BLR.

\bigskip
\line{\bf 4.3.1. Cross-Correlation of Integrated Line Fluxes\hfill}
\medskip

We now compare the results of the CCF lag time calculations
for model and data integrated lines in the K95 and the
C91 campaigns.  As with any numerical calculation involving data, there
is an error associated with the uncertainties in the input. The problem
in analyzing data lags and centroids is that there is no adequate
method in computing the errors (e.g., C91). Only if the peak of the CCF is
sharp, can the error estimates be made (Gaskell and Peterson 1987).
Errors presented here come from this estimate. More sophisticated
estimates of errors based on Monte-Carlo simulations also provide an
idea of the spread in lag times that will occur, but the results are
model dependent. The errors estimated this way are generally substantially larger than
those computed using the Gaskell and Peterson method, especially when the CCF is not
sharply peaked.

The behavior of NGC~5548 during the K95 campaign presents some
difficulties.  Ideally, it would have been advantageous to use only the
high-quality {\it HST} data to perform the CCF peak analysis.  To determine a
precise lag by utilizing the CCF, however, requires a significant change in
sign of the the slopes of the light curves. This happens
because, by the definition, a CCF becomes constant when either of the
light curves is a linear function of time.

During the last 39 days of the K95, when NGC~5548 was observed with the
{\it HST}, the C~IV line strength steadily increased.  It is far from being
clear, however, whether it had peaked by the end of the campaign. In
other words, the line curves do not turn significantly downward, as can
be seen in Fig.~5.  The apparent lack of the downturn is problematic
because during this time interval the emission light curve is approximately
linear, and, as a result the peak of the corresponding CCF is not well
defined, if only the {\it HST} data is used.  To overcome this difficulty, we
were forced to augment the {\it HST} portion of the K95 campaign with the
lower-quality integrated line flux of the {\it IUE} data.  This provided
coverage from JD 2,449,060 to JD 2,449,135.  During this time, the line
flux went down and then came back up again, so the response was reasonably
well sampled.  According to K95, four data points were suspect, i.e., JD
2,449,063, JD 2,449,077 JD 2,449,082 and JD 2,449,095, and have been
removed from our analysis.

The peak lag and centroids of the model and the data are presented at
the Table~2. For the K95 campaign, the model produced peak and centroid lags 
of 8.3  and 9.2 days. The Gaskell-Peterson error (hereafter GPE) is 
estimated at 0.6 days. The data peak lag was found to be 4.6 days with a 
GPE of 0.6 days; the centroid was 7.0 days. Thus, the model and data centroids
are reasonably close to each other, and peaks are within a factor of two.
However, the adequacy of GPE estimate of an overall error in computing the lag
is questionable and the real errors in data CCFs are probably larger,
making the differences between the model and data even less significant.
The simulations of White \& Peterson (1994) confirm our suspicions that in
the case of undersampling the response of the emission line, the GPE seriously
underestimates the uncertainty in the measured lag.

The C91 campaign involved only {\it IUE} data and as a result the
uncertainties in the flux values are larger.  In addition, the time
intervals between successive observations were larger as well.  As a
result, the GPE for all lag times presented below is between 2 to 3
days.  On the other hand, during the C91 campaign, the continuum rose
and fell three times and the C~IV line responded with three large
amplitude variation events of its own (see Fig.~5).  Thus the light
variations were well sampled.  We computed the lag times, $\tau_{Peak}$,
and centroids, $\tau_{Centroid}$ for
the three separate events and for the entire C91 campaign. The three
events were defined by the time intervals:  $t_1=$ JD 2,447,534.0 to JD
2,447,581.9, $t_2=$ JD 2,447,586.0 to JD 2,447,688.8, and $t_3=$ JD
2,447,692.7 to JD 2,447,737.8. The lag ranges for the various time
intervals are listed in Table~2, for both the model and the data.
The peak lags in the Table show that in all three events, the ordering
of the peak lag is the same.  The shortest is $t_3$ followed by $t_1$;
the longest is $t_2$. In events $t_1$ and $t_3$, the model and data lag
values fall within the GPE.  Also, the model peak lag determined
from the entire C91 campaign is consistent with the data. Because it too
falls within the GPE, this suggests that, on the average, the model is 
consistent with the data on a long term basis.  There is, however, a
significant difference between the lag times of the data and the model
in the second event.  The lag time $t_2$ for the data is over twice as
long as the model which is the reverse of what was observed in the K95
campaign.  In the K95 campaign the line light curve went
through a trough, whereas in the C91 campaign event $t_2$, the light
curve went through a peak.  One possible explanation is that, during a
sufficiently large increase in the continuum flux, an ionization front
pushes its way through the innermost clouds making these clouds
optically thin.  The bulk of the emission then shifts to higher radii
with longer lag times.  The opposite may occur in a low continuum state.
Fig.~3b shows that the continuum event which
gives rise to $t_2$ has the highest peak flux and goes through the
greatest change in flux.  In the K95 data set, the continuum was
initially approximately constant, then rapidly dropped and rose again, making 
the emission
light curve form a trough between the {\it IUE/HST} campaigns.  In principle, 
given this scenario, the ionization
front above the molecular disk would then migrate back towards the
continuum source and lag times would become shorter. Since, in our analysis,
the clouds are always optically thick, substantial changes in the time lags
are not expected with accompanying changes in the continuum.

\bigskip
\line{\bf 4.3.2. Cross-Correlation of Synthetic Line Segments\hfill}
\medskip

In this section we present the results of computing the lags for
the various line segments of the model for the C91 and the K95
campaigns.  We can not compare these to observations for the C91 campaign
because the individual line segment data is not reliable due to a low S/N. 
Also, in K95, the line segment 
data exists only during the {\it HST} portion of the campaign and suffers the same
problem as the integrated line does (see above). This means that the error
estimates by the Gaskell-Peterson method as well as from Monte Carlo simulations
will be of little help. In such a case, only the most general qualitative 
statements can be made about the actual lags and centroids based upon
these data.
 
In order to demonstrate our point that the data CCFs can not be
utilized for determining peak lags, we show as an example in Fig.~7a
the CCF of the observed MRW with the continuum. The broad flat top of the data
CCF is composed of two shallow bumps. One has a peak at 4 days and
another --- at 16 days. The bump at 16 days is slightly higher than the
one at 4 days, but the height difference is not significant, so there
is no clearly defined peak.  To an even higher degree, double and
multiple peaked flat top behavior is present in the rest of the data
CCFs as well, with the exception of the FBW CCF which shows only one
peak at 5 days, but the maximum correlation is weak.

The model lag times and centroids are presented in Table~3a. The first column 
indicates the wing segment. The next two colums show the lag time and centroid 
for the K95 campaign. The last two colums show the lag time and centroid for 
the C91 campaign. Table~3a reveals a sequence of responses of modeled line 
segments to the underlying continuum variation. The first segment to respond 
is the MRW, which is followed by the FBW, about a day later. Thus, in the model, 
the red side of the line responds first.  The total sequence, from first to 
last, is as follows: MRW, MBW, FBW, FRW, RC and BC, meaning that the wings 
respond first and the cores respond last.  Although there is 
a three-way tie between  the FBW, MBW and FRW in the peak lags of C91, the 
corresponding centroids follow the time sequence of the modeled response to 
the K95 continuum. 

The main outcome of theoretical analysis of the model is the prediction that 
different parts of the C~IV emission line respond in a particular order.
Though the causal agent of change in the model is the continuum, it is advantageous to compare the relative lags of individual line segments by 
direct cross-correlation of segment light curves. This accomplishes two goals. 
First, since noise is associated
with the continuum input, the error will propagate into each of the computed time
lags. By directly cross-correlating two line segments, we eliminate that
source of the noise. Second, it is possible that the
two line segments might not correlate with the continuum as well as they do
with one another.  In this case, relative time lags can be determined with 
a greater precision.

Lag times and centroids were calculated for a variety of line segment
pairs in the model. The results are presented in Table~3b.  In column
one, the segment pairs are listed in a particular order, where the
following convention is used. A positive lag indicates
that the second entry lags the first. In all seven comparisons of line
segments, there is a complete agreement between the lag and the
centroid entries of the K95 and C91 campaigns in the ordering of responses.

We note in passing, that when segment {\it vs.} segment cross-correlation is
attempted with the available data, the CCFs for line segment pairs show
single weak peaks superimposed on a wide plateau. Even worse are those
pairs that involve the FRW, because it is probably contaminated, and
those with the FBW, because its variations have low signal-to-noise. An
example of MBW {\it vs.} MRW is shown in Fig.~7b which exhibits the best CCF
obtained from the data.  This particular figure suggests that the MRW
responds before the MBW and is consistent with the claim in K95 that
the red wing of C~IV responds before the blue wing.  Although the far wings
respond here between the mid-wings and the cores, we caution that this
conclusion is model-dependent and is based on the local electron scattering
approximation.

To summarize the most general predictions of the model, (1) the MRW 
response preceeds the responses of all other C~IV line segments; (2) 
the mid-wings preceed the line cores.

\bigskip
\line{\bf 5. DISCUSSION\hfill}
\bigskip

The modified EBS model applied to NGC~5548 can successfully reproduce
the C~IV line profile variations detected by the {\it HST} observational
campaign and the line fluxes available from both campaigns. It is important to 
understand how the geometry and the
physics of this model BLR causes the emission line profiles to change
with time.  In particular, it must be explained how this model, which
has a strong outflow component, is able to produce the shortest lags on
the red side of the line.  In this section we address these issues and
also discuss the limitations of the present model. 

\bigskip
\line{\bf 5.1 The Model: Two-Dimensional Transfer Function\hfill}
\medskip

The two-dimensional transfer function (2DTF) provides a means of analyzing 
the velocity-resolved response of a system to a driving input, at relative
velocity $v$ and as a function of time $t$ (Blandford and McKee 1982; Welsh
and Horne 1991). The 2DTF, $\Psi(v,t)$, is related
to the luminosity of the driving continuum $L_c(t)$ and the line
profile strength $L_l(v,t)$ via
$$L_l(v,t)=\int\limits_{-\infty}^\infty dt'L_c(t')\Psi(v,t-t'). \eqno(10)$$ 
If a delta function continuum pulse is substituted into the eq.~(10),
then $L_l(v,t)=\Psi(v,t)$.  Thus, the 2DTF is the response of the system
to a delta function pulse.  By analyzing the 2DTF we may determine which parts
of the line profile respond first, which parts of the profile respond
last, and which parts of the profile respond with the greatest
strength.  The information gained  from the 2DTF may then be
interpreted in terms of a particular model.

Fig.~8a shows the two-dimensional transfer function (2DTF) for the
model. It was obtained by propagating a rectangular 0.1-day continuum
pulse. Since the NLR is stationary on the timescales discussed in this
paper, it was ignored in calculation of the 2DTF.  The
outer edge of the C~IV emitting region in the model is set by
obscuration of the central continuum source. At $30^o$ above the plane
of the disk, along the upper boundary of the BLR, the obscuration
effect is the weakest. At this angle, the central source, as viewed
from the outflowing clouds, is completely obscured at a distance of 24
light days. At an observer's aspect angle of $40^o$, a maximal time lag
of 28 days for this part of the C~IV emitting region will be measured.
The emission flux contribution from the region corresponding to this
delay, however, is negligible, so the bulk of the emission corresponds
to earlier response times.

As can be seen in Fig.~8a, the 2DTF consists of two parts, a
strong central region (the 2DTF core) between relative velocities $\pm 
6,000\ {\rm
km\ s^{-1}}$, and low flat wings on both sides of the central region.
The 2DTF core corresponds to the C~IV line core and the mid-wings. This core
grows rapidly for the first 4--5 days. The first response in the 2DTF is 
observed at around $\pm 7,000-7,500\ {\rm km\ s^{-1}}$. These emission `horns' 
arise at the contact points of the innermost possible paraboloid of constant
light travel time to the observer with the BLR (i.e., corresponding to the
minimal lag). In the next section we provide a detailed analysis of the
3D velocity field and of cloud emissivity function which fully explain the
time-dependent effects of the 2DTF uncovered here.

Both red and blue sides of the 2DTF core rapidly rise, but the red side
responds faster and more strongly than the blue side. This can be clearly 
seen in the contour density in Fig.~8a. This trend is repeated again during 
the decline, i.e., the red side of the 2DTF core drops rapidly in comparison 
with the blue side. The steepest decline occurs in the MRW. From MBW to BC
to RC, the decline becomes more shallow. The difference in the amplitude 
between the red and blue sides of the 2DTF core at early times is roughly 30\%.
As we show in section 5.2, this is due to excess of emitting gas with 
redshifted velocities at small $r$ over that with blueshifted velocities.

In the region of the low flat wings of the 2DTF, the emission grows in concert 
with its core, staying always low. The contour levels in the far wings, as
exhibited by Fig.~8a, do not exceed the 10\% level.  Most of the
emission in the far wings is due to electron scattering of the line core photons. 
Comparison of Figs.~8a and 8b shows that the
model 2DTF far wings are considerably reduced in the absence of scattering.  
This and the strength of the line core explain why they respond similarly 
to the line core.

The local scattering model for the far wing origin implemented in this
work has an important advantage of being conceptually and technically simple. 
However, it may be an oversimplification. The
non-local scattering model used in EBS, by calculating the
far wing contribution from the disk corona, is probably more realistic, 
especially because evidence points towards the Balmer lines far-wings being 
stationary in time in at least one case of Mrk~590 (Ferland, 
Korista and Peterson 1991; Peterson et al. 1993). Within the non-local scattering 
model, the far wings are 
formed at various radii from the central source, hence their response to continuum variability should be nearly obliterated.

\bigskip
\line{\bf 5.2 The Model: Projected Velocity Maps.\hfill}
\medskip

The velocity field of the EBS flow is a mix of rotation and outflow. 
At the lower boundary of the BLR, it is predominantly Keplerian, gradually 
making transition to outflow as the altitude above the disk plane is increased.
To understand the disposition and response of material emitting 
at a particular velocity, we dissect the model BLR by cutting it
with planes parallel to the accretion disk at different $z$.  We then map out
regions in each plane which correspond to the (line-of-sight) velocity 
intervals of the line segments, from the FRW to the FBW. This enables us to follow the 
changes in the spatial distribution of velocity-bounded intervals
as the altitude is increased. 

In the model presented in this paper, an observer views the BLR looking 
down at the disk, at an aspect angle of $40^o$ with the $z$-axis (Fig.~1).
In the absence of electron scattering contribution to the C~IV line shape,
it is possible to identify regions in the BLR which contribute to the particular line
segments. In other words, in this case, the shape of the modeled line profile 
corresponds to the actual distribution of emission in the velocity space of
the line-of-sight velocities. Fortunately, as shown by Figs.~8a and 8b, the 
core and the mid-wings of C~IV line (or the 2DTF core) have
negligible contribution from the electron-scattered light. Therefore, the behavior of the 
C~IV core and mid-wings (or the 2DTF core) can be understood 
by analyzing the spatial distribution of relative velocities in the BLR model.
At the same time, the far-wings of the C~IV line (and the wings of the 2DTF) are 
heavily dominated by electron scattering, which makes them unsuitable for 
this type of analysis. In this section, we analyze the C~IV emission distribution
based on the 3D kinematics of the outflow from the disk neglecting the contribution 
from the electron scattering. Only unscattered light in the FRW and the FBW is
accounted for.
  
Fig.~9 presents eight two-dimensional cross-sections of the BLR at
different altitudes above the disk plane. The vertical axis is
the cylindrical radius, $r$, measured in light days.  The horizontal axis is
the azimuthal angle measured in degrees. The gray scale in Fig.~9
demarks the regions which have line-of-sight velocities that
correspond to the velocity intervals of the six line segments defined
in section 4.2.2. The shades of gray, from darkest to lightest, correspond to the 
FRW-to-FBW sequence. The redshifted rotational velocities are between $90^o-180^o$, and
the blueshifted ones between $180^o-270^o$. All areas shaded black in Fig.~9 
represent spatial regions 
which do not contribute any emission to the line profile because of the cloud emission 
anisotropy. The observer is at an azimuthal angle of $0^o$. Angles smaller than $90^o$
and larger than $270^o$ are not shown, as they would be shaded entirely black,
because this part of the BLR is not observable. The two horizontal lines in  
Figs.~9a--h provide the spatial extention (the inner and outer cylindrical boundaries)
of the emitting BLR at each altitude. The observed C~IV emission is strictly limited 
to lie within these boundaries. The cross-hatched region above the upper
horizontal line represents a region of the outflow that does not emit C~IV
because of the cloud obscuration, here $\tau_{obs} > 1$. 
The cross-hatched region below the lower horizontal line is over-ionized, i.e.,
it lies within the inner boundary of the BLR, and no C~IV photons are generated 
here. Because Fig.~9 accounts for the {\it unscattered} emission 
component only, it must be considered in conjunction with the Fig.~8b (the 
unscattered 2DTF).

Inspection of the Figs.~1 and 9 reveals that the C~IV emitting regions at higher $z$ 
cross-sections are also located at greater distances from the continuum source than 
the lower ones. This means that the possible response times for an emitting cloud 
become longer as the altitude of the cross-section is increased. 
What complicates the visualization of the 3D velocity field is that there are
two components to the outflow velocity which must be accounted for, i.e., along the 
$z$ and $r$-axes.

Close to the disk surface, the velocity field is dominated by Keplerian rotation
with a small addition of $r$ and $z$-components of the outflow (Fig.~9a).
Rotation is responsible for the overall symmetry, as seen from the distribution of gray-shaded areas around the azimuthal angle of $180^o$, meaning that the line
profile must be more or less symmetric in the core. A close inspection of Fig.~9a
reveals that the area corresponding to the MRW is larger than the area of the MBW.
This prevalence of redshifted C~IV emission in the mid-wing is the signature 
of the $r$-component of the outflow. The clouds that contribute to the MRW at
this low altitude lie at small $r$ and have maximal allowable line-of-sight 
velocities directed away from the observer. Their emission contribution is substantial 
because of two reasons. Firstly, being at small $r$, they are subject to a large 
incident flux. Secondly, because of the emission anisotropy, clouds moving away from 
the observer expose their illuminated surfaces. Thus the redshifted emission is 
particularly bright in the 2DTF.

The signature of the $z$-component of the outflow velocity at small $r$ (corresponding 
to small lag times) in the 2DTF can be observed at the first contact points, 
occuring about 0.1 light days above the disk, near $\pm7,000-7,500\ 
{\rm km\ s^{-1}}$, (Fig.~8b; section~5.1). The
corresponding emission `horns' in the 2DTF are not symmetrically placed in
Fig.~8b, both being slightly blueshifted compared to their projected Keplerian
velocities at the first-contact points. The origin of these `horns' was
discussed in section~5.1 in the context of the time lags: these are the contact 
points (intersections) of the innermost possible paraboloid of constant light 
travel time to the observer with the C~IV emitting gas. The emission `horns' appear
blueshifted because the first contact points are near the disk plane
at almost right 
angles with respect to the observer's line-of-sight. The projection of the radial 
outflow component onto the observer's line-of-sight nearly vanishes here. 
Thus, the blueshifted $z$-component in the outflow is emphasized.  The
observed emission at the first contact points is small, because the clouds, observed 
here at almost right angles, have their C~IV emissivity heavily suppressed by the 
emission anisotropy.

As the cross-section altitude increases, the velocity field becomes distorted towards 
the blue (Figs.~9b--h). This is a direct
consequence of the growing outflow velocity in the $z$-direction
superimposed on the rotational velocity component. The $r$-component of the outflow
velocity decreases with height. If it were not
for the presence of an opaque underlying accretion disk, the
emission would be distributed symmetrically in velocity space.

The altitude dependence of the velocity field is seen most clearly in the
MRW, which turns into an `island' at 1 light day above the disk. By
2 light days above the disk, the MRW has disappeared completely.
Thus, of the four velocity intervals that make up the 2DTF core, the
response of the MRW is more `tightly bound' than the other line segments
(these are still present, even at a cross-section altitude of 8 light
days). Consequently, the MRW responds before the RC, BC or the MBW.
The unscattered FRW and FBW in Fig.~9 are bound even more
tightly than the MBW, and as a result respond even faster than the MRW.
However, their intensity is so low that, when the scattered light is superimposed
on it, the response is dominated by the (RC and BC) scattered light.  
Inspection of Fig.~8a and 8b shows that the far wings peak later on, if 
the electron scattering is present. 

As for the remaining three regions which make up the 2DTF core, the RC, BC and 
MBW, a migration of the (gray shaded) boundary between the RC and the BC and 
also the BC and the MBW occurs as the cross-section altitude is increased. In each 
cross-section in Fig.~9, the mean radial position of the RC region is larger 
than that of the BC, and the mean position of the BC is larger than that of the MBW.  
The MBW hence will respond before the BC, and the BC is expected to respond
before the RC. This explains the order of the responses in the 2DTF core, i.e.,  
the MRW responds first followed by the MBW, the BC, and finally the RC.

%\vfill\eject
\bigskip
\line{\bf 5.3. The Model and the Data: One-Dimensional Transfer Function\hfill}
\medskip

The one-dimensional transfer function, 1DTF, shows the temporal emission
response of the overall BLR. The
mean line profile, which is the time average of the line profiles, on
the other hand, does not provide timing information on the BLR, but
shows the steady state line profile.  In this section, we analyze both
of these functions calculated by the model, in an effort to compare them
with those derived from the data.

We obtain the 1DTF by integrating the 2DTF along the line-of-sight velocity axis, 
(Fig.~10). The resulting 1DTF shows a peak 
at 4 to 5 days and trails off to zero intensity at 28 days. The centroid of 
the model 1DTF is at 8.6 days. 

The behavior of the model 1DTF in Fig.~10 can be compared to transfer functions
derived from the monitoring of NGC~5548 (Krolik et al. 1991; Peterson 1993; 
Wanders et al. 1995; 
Done and Krolik 1996) by a variety of inversion methods used to extract the 1DTF 
from the data. In Wanders et al. (1995), who used the K95 data,
the 1DTF declines toward zero intensity at large lags (25 days), similar to our model.
At small lags, the
data 1DTF is compatible with a wide plateau or even a slight depression within
5 days, although Wanders et al. preferred a monotonically rising function 
toward the zero lag. Our model 1DTF dives to zero intensity at 1 day, due to the 
cloud emission anisotropy.  

It is important to understand that the differences between the data and the model 
1DTFs at small lags could, at least in part, be due to resolution of the data 1DTF, 
i.e., the inversion 
process which recovers the data 1DTF is subject to the temporal resolution of the
discrete data set and the signal-to-noise ratio of the light curves.
In order to compare the model and data 1DTFs, we
follow the prescription of Wanders et al. in which the model 1DTF
is first degraded by convolving it with a Gaussian having a 3.8 day
standard deviation (dashed line in Fig.~10).  We find that, as before,
the resulting 1DTF trails off toward zero near 28 days, but the amplitude at zero lag
is at 0.52. The peak and the centroid
are shifted to about 7 and 9.2 days, which are near the shoulder of 
the data 1DTF in Wanders et al. In principle, the addition of a rapidly
responding, more isotropic emission component to the model should introduce response
at small lags and cause the present peak of the 1DTF to appear as a shoulder or as 
a secondary peak. We did not pursue this avenue, however.

Done and Krolik (1996) also used the K95 campaign to derive the 1DTFs. Whether a 
shoulder is present in their 1DTFs depends on the choice of optimization parameters 
and whether they use only the {\it HST} or the combined {\it HST/IUE} data sets, 
i.e., a
shoulder is present when only the {\it HST} data is used. In both cases the
peak of the 1DTF occurs at zero lag. However, temporal structures of less than
2 days are not resolved, so it is uncertain if the peak is at zero days or at a 
slightly later time.

Peterson (1993) presents response functions for a variety of emission lines, 
including C~IV. The 1DTFs were extracted from the C91 campaign and are an updated 
version of the results published in Krolik et al. (1991). The 1DTF of C~IV
is peaked at zero lag and descends to a low plateau at 25 days.  The amplitude 
drops off to zero at 100 days.  There is a small secondary peak between 150
and 200 days, but it is thought to be an artifact due to correlations
between the three separate events of the C91 campaign. Our model is consistent
with the response function for C~IV in Peterson (1993).

The above discussion focuses on the difference between our model and
the data-derived response functions at small lags as being possibly due
to the temporal resolution of the inversion methods.  Residual
differences between the model and the data 1DTFs at small lags can also be 
understood if low column density clouds are present in the flow.
Such clouds would emit a fraction of their C~IV line flux from the backsides,
which is neglected in the present work.
Response at zero (or small) lags, as indicated by the data 1DTF, means that there 
is emitting material along or close to the observer's line-of-sight.

\bigskip
\line{\bf 5.4. The Mean Line Profile\hfill}
\medskip

Fig.~11 presents the mean line profiles of the data and the model for
the K95 campaign.  The mean line profiles were generated by averaging
the line profiles for the data and the model during the {\it HST} portion of
the K95 campaign.  The model mean line profile is also generated by
integrating the 2DTF along the time axis.  When the contribution of the
NLR is removed from the model and compared to the time integration of the
2DTF, we find that they are identical, providing an independent 
check on the calculation of the 2DTF.

The essential features of the mean data profile are reproduced by the
mean model profile. The differences are partly due to the absorption
which is present in the blue part of the data line profile (i.e., in the
BC) and the contamination with He~II and possible Fe~II emissions in
the red wing.

In the RC and between the BC and the MBW, the model shows some excess
of emission above the data. Additional emission can be explained in
terms of our assumption that all the clouds are optically thick. 
Being efficient emitters, optically thick
clouds produce more emission compared to the optically thin clouds,
and our assumption is especially vulnerable at the inner and upper boundaries
of the calculated BLR model. Realistically, optically thin clouds
are likely to be found here, especially when the continuum is bright. 
Note, that the excess emission is not symmetrically placed with respect
to the zero relative velocity shift. The red excess lies within the RC
whereas the blue excess straddles the BC and the MBW. This argues
against this excess emission originating close to the disk because the
velocity field there is dominated by the Keplerian rotation. In this
case, the excess emission would be symmetrically placed with respect to
the zero relative frequency shift. Hence, the additional emission must come
from the upper portion of the BLR, where the outflow dominates and where
much of the low column density, optically thin (dustless) material is expected to be 
found. Exactly
this can be seen in the dissections presented in Fig.~9, where, at the
distance of more than two light days above the disk, the BLR contains
only RC-, BC- and MBW-emitting material.

The model FBW does not fit the data well. Between
the intrinsic and Galactic C~IV absorption features the data mean line profile can 
be modeled with a single power law fit, which breaks down between $-5,400\ 
{\rm km\ s^{-1}}$ and $-9,600\ {\rm km\ s^{-1}}$.  This is in contrast with
the red side which can be fit with a single power law, all the way into the core, 
up to the NLR part. The above velocity interval is perhaps contaminated by Si~II 
emission.  However, the degree of contamination to the C~IV emission is uncertain 
and so the modeling there represents a compromise between extreme cases.

The model mean line profile also shows a slight $\sim10$\% deficit of
emission in the FRW.  Part of the
FRW is accounted for, because contamination corrections for He~II and
Fe~II have not been subtracted off the mean data line profile (note
that coarse corrections for these elements have been applied to the line
segments mentioned earlier in section 4.2.1).  In addition, the broad He~II
emission varies on a shorter time scale than the genuine C~IV FRW
(as clearly seen in the Fig.~6d for the FRW prior to JD 2,449,115). Its
modeling is outside the scope of this paper. 

\bigskip
\line{\bf 5.5. Limitations of the Model\hfill}
\medskip

In addition to the simplification of the emission anisotropy, there are other 
issues that need to be discussed. First, the EBS MHD model is a self-similar 
model providing scaling relationships for hydromagnetic flow and cloud parameters.
We avoid a self-consistent calculation of C~IV volume emissivity due to its
complexity. Instead, we use a mix of cloud properties which is qualitatively
consistent with the properties prescribed by the EBS model. The
CLOUDY-calculated emissivity function reflects our assumption that the
clouds are optically thick throughout the BLR. This necessarily ignores
the thick-to-thin transition that some clouds will experience in
response to the increase in the central continuum. Such a transition
would ultimately lead to over-ionized clouds, to decreased cloud C~IV
emissivity and to a change in the cloud emission anisotropy (partly due to the
dust destruction), modifying the response times of the BLR to continuum variations.

Second, the cloud emissivity function depends on the (unknown) ionizing continuum flux
the clouds are subject to. Inspection of the continuum windows in K95 (see 
Fig.~6 of K95) reveals that at shorter wavelengths the continuum variations 
become larger and the peaks and valleys of the light curves become sharper.
If this trend continues to even shorter wavelengths, the relevant ionizing 
continuum variations could be significantly different in character from what is 
being used here.

Third, the overall geometry of the C~IV emitting region was assumed to be
time-independent. This certainly is an approximation, as we
expect that the upper and outer boundaries of the emitting region to move in 
response to the continuum variations. In particular, the opening angle of the
wind which is fixed now at $120^o$ should vary because clouds close to the upper
boundary may become over-ionized and perhaps destroyed. The outer boundary of the C~IV
emitting region, set by the condition $\tau_{obs}=1$, will move as well because
the obscuration will change with the variations in the ionizing continuum. 

A related issue arises from the approximation that has been made in the
radiation transfer through the cloud system.  As mentioned in section
3, clouds in the model have sufficient optical depth to obscure the
central continuum source in some directions. The purely geometrical
obscuration used assures that the clouds either see all or a fraction
of the original continuum whose shape was unaltered. A more realistic
model should account for the continuum reprocessing by the clouds and
in particular the changing continuum shape. The latter will affect the
cloud spectrum and response to the continuum input. Accounting for this effect,
however, requires the full solution of the radiative transfer problem
which is outside the scope of this paper.

Finally, the model presented in this paper was not fully optimized, so, in
principle, better fits to line profiles and measured time lags are
possible. However, we have shown that the hydromagnetic wind model
is capable of reproducing the main features observed in both campaigns.
A more detailed analysis is the subject of a future work.

\bigskip
\line{\bf 6. CONCLUSIONS\hfill}
\bigskip
  
We attempt to infer the dynamical state of the BLR in the Seyfert~1 galaxy NGC~5548
by analyzing results of the long term observations and by providing a critical 
comparison with the predictions of a hydromagnetically-driven outflow model 
from an optically-thick accretion disk. Our modeling is based on the 
EBS model, subject to a number of modifications, i.e., cloud emission anisotropy 
and cloud-cloud obscuration. We have shown that a model of a rotating 
magnetized wind from a disk is compatible with the observed line profiles, and 
the BLR line and continuum variability monitored in C91 and K95 campaigns. 

We have demonstrated that a time series of synthetic C~IV emission lines can
be fit to the {\it HST}-observed line profiles reasonably well, using the observed UV 
continuum light curve as the driver and {\it without adjusting the model parameters}.
Our best model for the observed C~IV emission in NGC~5548 consists of a disk-like 
region extending $\pm30^o$ in the polar directions from the equatorial plane and 
between 1 lt-d and 24 lt-d in the radial direction. Our model is also highly 
anisotropic because radiation is essentially backscattered. The C~IV-emitting part of 
the wind is bounded by the molecular wind at larger radii, depending on the
altitude above the disk. We expect the low ionization lines, like Mg~II, to 
come from the region where the central UV-soft-X-ray continuum is attenuated, 
i.e., from $\tau_{obs}>1$. The best fit to the observed C~IV line profiles also
constrained the mass of the central black hole to $\sim 3\times 10^7\ 
{\rm M_\odot}$. Surprisingly, the variable integrated line flux calculated within 
the model also fits reasonably well  
that from the {\it IUE} (C91) campaign which happened four years prior to the 
{\it HST} campaign, again without any parameter adjustments or flux renormalization.

We have divided the synthetic C~IV line profile into 6 segments 
corresponding to the blue and red cores, and mid and far-wings, and calculated the 
cross-correlations between different line segments as well as between the line 
segments and the continuum. The model predicts that the mid-red wing response 
preceeds the responses of all other C~IV line segments, and that the wings respond 
before the line cores. Given that no adequate method in computing the errors for 
data lags and centroids exists in the literature, the data cross-correlation 
function provides results which appear inconclusive, making any direct comparison 
with the model premature. The only statement we make in comparing the data lags and 
centroids with the model, is that they are compatible.

In order to understand the model 
response, the underlying geometry and the C~IV emission distribution in
the velocity field in the model BLR, we have computed the one-dimensional and 
two-dimensional transfer functions. We find that, due to the cloud emission 
anisotropy, an observer at a typical aspect angle of $40^o$ to the rotation axis 
receives the main contribution 
to the C~IV line coming from the backside of the wind. Consequently, at the base of 
the wind, the emitting gas kinematics is dominated by a rotation plus an
outflow velocity projected away from the observer. It is the shape of the streamlines 
in the $rz$-plane (parabolas), which leads to a switch from the redshifted outflow 
velocities to the blueshifted ones at some height above the disk. Initially, the red 
part of the C~IV line grows faster than the blue one. The 
steepest decline in the redshifted emission appears in the mid-red wing which 
turns over first. This predicts a differential response across the C~IV profile, 
in particular, a redward asymmetric response. Unfortunately, no unambigous
differential lags have been measured across the C~IV line during the {\it HST} part of the
K95 campaign, and so the resolution of this problem awaits further observations.

The idea that the outer parts of accretion disks in AGNs can be molecular
is supported by detection of such gas in the inner regions of AGNs (Scoville
et al. 1989) and by the possibility that all AGNs are fueled by a dusty and 
molecular galactic ISM (Shlosman, Frank and Begelman 1989). If this material 
can be raised above the disk, it can easily
obscure the central source by intercepting and reprocessing a large fraction of the
UV-soft-X-ray luminosity into the infrared. Sy~2 galaxies, with NGC~1068 as their 
prototype, fall within this category of objects having intervening column
densities in excess of $10^{25}\ {\rm cm^{-2}}$ (Krolik and Begelman 1988; Mulchaey,
Mushotzky and Weaver 1992). Hard X-rays will be least affected and hence may serve
as diagnostic of AGNs viewed through a large column of obscuring material almost
supressing the BLRs, e.g., in the so-called narrow-line X-ray galaxies (e.g., Lawrence
1991). The possibility that the obscuration comes from the high-column density 
material in the hydromagnetic wind was used by EBS to explain LILs and the broad
absorption-line quasars, and was incorporated by K\"onigl and Kartje (1994) into
the unification scheme of the AGNs. Adopting the wind origin of the obscuration
resolves the problem of vertical support for molecular tori in Seyfert galaxies,
described by Krolik and Begelman (1988).

In this work we have largely neglected the dynamical effects of radiation pressure
from the central continuum on the magnetized wind. Although this effect cannot be 
ignored in brightest AGNs, as shown by de Kool and Begelman (1995) for the case of 
broad absorption-line quasars, it might not dominate the cloud dynamics in lower 
luminosity Seyfert galaxies. This is demonstrated by an apparent survival of the 
molecular torus in NGC~1068 which is geometrically thick and has an opening angle 
which is only slightly smaller than the angle of $120^o$ inferred here for NGC~5548. 
Along with K\"onigl and Kartje, we do expect that central radiation flux will 
control the opening angle of the wind, which is consistent with observation that
the thickness of molecular tori in Seyfert galaxies seem to decrease with 
increasing central luminosities (Lawrence 1991; Miller 1994). It is unclear whether
the wind opening angle can respond to short-time variability of the central continuum,
such as observed in NGC~4151 which changed its appearance from Sy~1 
to Sy~1.9, as the continuum luminosity decreased (Clavel et al. 1990; Ulrich et 
al. 1991).

In summary, this model  provides an
attempt to match a dynamical model to the monitoring campaigns of NGC~5548. 
We have avoided a systematic 
parameter search over the entire phase space available for modeling, so the model 
presented here is neither guaranteed to be unique or to provide an optimal fit to 
the data. However, we conclude, based on our analysis, that a rotating outflow from an 
accretion disk provides a viable model for the BLR in NGC~5548, in particular, and 
for other AGNs, in general.

\smallskip
ACKNOWLEDGEMENTS. We thank Mitch Begelman, Chris Done, Matthias Dietrich, 
Julian Krolik and Brad Peterson for valuable discussions, and the anonymous
referee for helpful comments. We are grateful to 
Gary Ferland for assistance with the recent version of CLOUDY. This work 
was supported in part by NASA grants WKU-521751-94-08 and 521752-95-06 
and NAGW-3839 (IS), NAG-3223 (KTK), and NSF grants AST-9223370 and AST-9529170 (RDB).
\vfill\eject

\centerline{\bf Table~1a}
\centerline{\bf Model Parameters}
\medskip

$\theta_0$ \hskip 1.449 in     $20^o$

$\lambda$ \hskip 1.498 in      5

$b$ \hskip 1.505 in            1.5

$\alpha$ \hskip 1.485 in        2.5

$T_e$ \hskip 1.434 in         $10^6$ K

$f_s$ \hskip 1.462 in          0.45

$i$ \hskip 1.51 in           $40^o$

$\psi_l$ \hskip 1.43 in      $3^o$

$\psi_u$ \hskip 1.407 in      $30^o$

$M_{BH}$ \hskip 1.245 in        $3\times 10^7\ {\rm M_\odot}$                    

\bigskip\bigskip
\centerline{\bf Table~1b}
\centerline{\bf Definitions of C~IV Line Segments in Velocity Space}
\medskip

Line Segment\hskip 1.55 in Velocity Interval

\hskip 2.67 in            (${\rm km\ s^{-1}}$)

Far-Blue Wing (FBW)\hskip 1.05 in  -9,000 to -6,000           

Mid-Blue Wing (MBW)\hskip .985 in  -6,000 to -3,000

Blue Core (BC)\hskip 1.48 in       -3,000 to 0.0

Red Core (RC)\hskip 1.56 in         0.0 to 3,000

Mid-Red Wing (MRW)\hskip 1.06 in    3,000 to 6,000

Far-Red Wing (FRW)\hskip 1.139 in    6,000 to 9,000

\medskip
\vfill\eject

\centerline{\bf Table~2}
\centerline{\bf Time Lags for Integrated Line: Data and Model$^a$}
\medskip

time interval\hskip 1.5 in    data \hskip 1.6 in              model

and campaign\hskip .789 in $\Delta t_{Peak}$\hskip .323 in $\Delta t_{Centroid}^b$ 
\hskip .399 in $\Delta t_{Peak}$ \hskip .362 in $\Delta t_{Centroid}^b$ 

\hskip 1.637 in (days)\hskip .594 in (days)\hskip .68 in (days)\hskip .654 in (days)

integrated line (C91)\hskip .6 in 8.0\hskip .679 in 11.7\hskip .85 in 7.0\hskip .78 in 8.9

$t_1 (C91)$\hskip 1.4 in 8.0\hskip .727 in 9.9\hskip .87 in 9.4\hskip .78 in 9.5

$t_2$ (C91)\hskip 1.300 in 24.0\hskip .6427 in 20.1\hskip .8313 in 10.0\hskip .764 in 7.6

$t_3$ (C91)\hskip 1.352 in 4.0\hskip .727 in 5.9\hskip .873 in 7.0\hskip .780 in 7.3

integrated line (K95)\hskip .5815 in 4.6\hskip .725 in 7.0\hskip .866 in 8.3\hskip .781 in 9.2

\noindent $.^a$ refer to section 4.3.1 for a discussion on the uncertainties in the lags.

\noindent $.^b$ centroid measured at the 80\% level.                

\vfill\eject

\centerline{\bf Table~3a}
\centerline{\bf Model Time Lags for C~IV Line Segments vs Continuum}
\medskip

line segment\hskip .38 in $\Delta t_{Peak}$\hskip .384 in $\Delta t_{Centroid}^a$\hskip 
.359 in $\Delta t_{Peak}$\hskip .343 in $\Delta t_{Centroid}^a$  

\hskip 1.128 in (days)\hskip .663 in (days)\hskip .647 in (days)\hskip .548 in (days)

\hskip 1.2 in      K95\hskip .72 in  K95\hskip .73 in C91\hskip .65 in C91 

FRW\hskip .9 in  7.5\hskip .83 in   8.7\hskip .77 in  6.6\hskip .67 in 8.4          

MRW\hskip .86 in  5.5\hskip .83 in   6.8\hskip .76 in  5.8\hskip .673 in 6.2

RC\hskip 1.01 in   8.4\hskip .84 in   9.5\hskip .764 in  9.6\hskip .667 in 9.2
 
BC\hskip 1.015 in   9.4\hskip .761 in   10.3\hskip .7469 in 9.9\hskip .6296 in 10.5
 
MBW\hskip .834 in  6.6\hskip .822 in   8.2\hskip .7754 in 6.6\hskip .677 in 7.8

FBW\hskip .864 in  7.4\hskip .823 in   8.5\hskip .785 in 6.6\hskip .667 in 8.2

Integrated\hskip .572 in 8.3\hskip .813 in  9.2\hskip .792 in  7.0\hskip .667 in 8.9

$.^a$ centroid measured at the 80\% level.    

\bigskip\bigskip
\centerline{\bf Table~3b}
\centerline{\bf Model Time Lags for C~IV Line Segments vs Line Segments}
\medskip

lines compared\hskip .453 in $\Delta t_{Peak}$\hskip .466 in $\Delta t_{Centroid}^a$\hskip 
.497 in $\Delta t_{Peak}$\hskip .503 in $\Delta t_{Centroid}^a$

\hskip 1.34 in (days)\hskip .875 in (days)\hskip .700 in (days)\hskip .730 in (days)

\hskip 1.422 in            K95\hskip .938 in K95\hskip .852 in    C91\hskip .851 in    C91

BC vs. RC\hskip .78 in -0.8\hskip .878 in -0.66\hskip .86 in   -0.8\hskip .864 in -1.03

MBW vs. MRW\hskip .46 in -1.1\hskip .9 in -1.01\hskip .854 in  -1.1\hskip .841 in -1.25

FBW vs. FRW\hskip .5715 in  0.2\hskip .935 in  0.43\hskip .907 in  0.2\hskip .89 in  0.14

MRW vs. FRW\hskip .549 in  1.5\hskip .931 in  2.07\hskip .907 in  1.7\hskip .887 in  2.02

RC vs. MRW\hskip .6325 in  -2.5\hskip .888 in -2.46\hskip .870 in -2.6\hskip .839 in -3.01

FBW vs. MBW\hskip .4875 in -0.2\hskip .888 in -0.15\hskip .8725 in -0.2\hskip .823 in -0.54

MBW vs. BC\hskip .657 in   2.2\hskip .934 in  2.58\hskip .9085 in  2.3\hskip .875 in  2.67

$.^a$ centroid measured at the 80\% level.    

\vfill\eject

\bigskip
\line{\bf REFERENCES\hfill}
\bigskip

\ref{Baldwin, J.A., Ferland, G.J., Korista, K.T. \& Verner, D. 1995, \ApJ{455}{L119} }

\ref{Blandford, R.D. \& McKee, C.F. 1982, \ApJ{255}{419} }

\ref{Blandford, R.D. \& Payne, D.G. 1982, MNRAS, 199, 883 }

\ref{Cassidy, I. \& Raine, D.J. 1996, MNRAS, submitted}

\ref{Chiang, J., Murray, N. 1996, ApJ, submitted}

\ref{Clavel, J. et al. 1990, MNRAS, 246, 668}

\ref{Clavel, J. et al. 1991, \ApJ{366}{64}\ (C91) }

\ref{Collin-Souffrin, S. \& Lasota, J.-P. 1989, PASP, 100, 1041}

\ref{Corbin, M.R. 1990, \ApJ{357}{346} }

\ref{Crenshaw, D.M. \& Blackwell, J.H. 1990, \ApJ{358}{L37} }

\ref{de Kool, M. \& Begelman, M.C. 1995, \ApJ{455}{448} }

\ref{Dietrich, M. et al. 1993, \ApJ{408}{416} }

\ref{Done, C. \& Krolik, J.H. 1996, \ApJ{463}{144} }

\ref{Emmering, R.T., Blandford, R.D. \& Shlosman, I. 1992, \ApJ{385}{460}\ (EBS) }

\ref{Espey, B.R., Carswell, R.F., Bailey, J.A., Smith, M.G. \& 
	Ward, M.J. 1989, \ApJ{342}{666} }

\ref{Fabian, A.C., Guilbert, P., Arnaud, K., Shafer, R., Tenant, A. \& Ward, M.
        1986, MNRAS, 280, 574}

\ref{Ferland, G.J. 1994, Hazy: A Brief Introduction to CLOUDY, University of
        Kentucky, Department of Physics \& Astronomy Internal Report}

\ref{Ferland, G.J., Korista, K.T. \& Peterson, B.M. 1990, \ApJ{363}{L21} }

\ref{Ferland, G.J., Peterson, B.M., Horne, K., Welsh, W.F. \& Nahar, S.N.
        1992, \ApJ{387}{95} }

\ref{Gaskell, C.M. 1982, \ApJ{263}{79} }

\ref{Gaskell, S.M. \& Peterson, B.M. 1987, \ApJS{65}{1} }

\ref{Gondhalekar, P.M., Horne, K. \& Peterson, B.M. (eds.) 1994, Reverberation
         Mapping of the Broad-Line Region in AGNs (San Francisco: PASP Vol. 69).}

\ref{Horne, K., Welsh, W.F. \& Peterson, B.M. 1991, \ApJ{367}{L5} }

\ref{K\"onigl, A. \& Kartje, J.F. 1994, \ApJ{434}{446} }

\ref{Koratkar, A.P. \& Gaskell, C.M. 1991, \ApJ{345}{637} }

\ref{Korista K.T., et al. 1995, \ApJS{97}{285}\ (K95)}

\ref{Krolik, K.T. \& Begelman, M.C. 1988, \ApJ{329}{702} }

\ref{Krolik, K.T. \& London, R.A. 1983, \ApJ{267}{18} }

\ref{Krolik, J.H., Horne, K., Kallman, T.R., Malkan, M.A., Edelson, R.A.
       1991, \ApJ{371}{541} }

\ref{Kwan, J. \& Carrol, T.J. 1982, \ApJ{261}{25} }

\ref{Lawrence, A. 1991, MNRAS, 252, 586}

\ref{Maoz, D., Netzer, H., Mazeh, T., Beck, S., Almoznino, E., Leibowitz, E.,
      Brosch, N., Mendelson, H. \& Laor, A. 1990, \ApJ{351}{75}

\ref{Maoz, D. et al. 1993, \ApJ{404}{576} }

\ref{Matthews, W.G. 1982, \ApJ{258}{425} }

\ref{Matthews, W.G. 1986, \ApJ{305}{187} }

\ref{Matthews, W.G. \& Capriotti, E.R. 1985, in Proc. Santa Cruz Workshop on
       Astrophysics of Active Galaxies and QSOs, ed. J.S. Miller (Univ.
       Science Books), p.~185.}

\ref{Matthews, W.G. \& Ferland, G.J. 1987, \ApJ{258}{425} }

\ref{Miller, J.S. 1994, in Proc. I Stromlo Symp. on Physics of Active Galaxies,
        eds. Bicknell, V. et al. (San Francisco: PASP Vol. 54), p.~149.}

\ref{Mulchaey,J.S., Mushotzky, R.F. \& Weaver, K.A. 1992, \ApJ{390}{L69} }

\ref{Murray, N., Chiang, J., Grossman, S. \& Voit, G.M. 1996, ApJ, in press}

\ref{Netzer, H. 1990, in Active Galactic Nuclei, eds. T.J.-L. Courvoisier \& 
           M. Mayor (Berlin: Springer-Verlag).}

\ref{Osterbrock, D.E. 1996, in The Violent Universe: Key Problems in Astronomy \&
           Astrophysics, eds. G. M\"unch et al. (Cambridge: Cambridhe Univ. Press), 
           in press.}

\ref{P\'erez, E., Robinson, A., \& de la Fuente, L. 1992a, MNRAS, 255, 502}

\ref{P\'erez, E., Robinson, A., \& de la Fuente, L. 1992b, MNRAS, 256, 110}

\ref{Peterson, B.M. 1987, \ApJ{312}{79} }

\ref{Peterson, B.M. 1993, PASP, 105, 247}

\ref{Peterson, B.M., Korista, K.T. \& Cota, S.A. 1987, \ApJ{312}{L1} }

\ref{Peterson, B.M., Reichert, G.A., Korista, K.T. \& Wagner, R.M. 1990, \ApJ{352}{68} }

\ref{Peterson, B.M. et al. 1991, \ApJ{368}{119} }

\ref{Peterson, B.M. et al. 1993, \ApJ{402}{469} }

\ref{Phinney, E.S. 1989, in Theory of Accretion Disks, ed. F. Meyer et al.
     (Dordrecht: Kluwer), p.~457.}

\ref{Rees, M.J. 1987, MNRAS, 228, 47P}

\ref{Robinson, A. 1995, MNRAS, 272, 647}

\ref{Rosenblatt, E. \& Malkan, M.A. 1990, \ApJ{350}{132} }

\ref{Sanders, D.B., Phinney, E.S., Neugebauer, G., Soifer, B.T. \& Matthews, K.
      1989, \ApJ{347}{29} }

\ref{Scoville, N.J., Sanders, D.B., Sargent, A.I., Soifer, B.T. \& Tinney, C.G.
      1989, \ApJ{345}{L25} }

\ref{Shields, G.A. 1978, in Pittsburgh Conf. on BL Lac Objects, ed. A.M. Wolfe
       (Pittsburgh: Univ. of Pittsburgh Press), p.~257.}

\ref{Shlosman, I. \& Begelman, M.C. 1989, \ApJ{341}{685} }

\ref{Shlosman, I., Frank, J. \& Begelman, M.C. 1989, Nature, 338, 45}

\ref{Shlosman, I., Vitello, P. \& Shaviv, G. 1985, \ApJ{294}{96} }

\ref{Stirpe, G.M., de Bruyn \& van Groningen, 1988, \AsA{200}{9} }

\ref{Ulrich, M.-H. et al. 1991, \ApJ{382}{483} }

\ref{Walter, R. et al. 1994, \AsA{285}{119} }

\ref{Wamsteker, W. et al. 1990, \ApJ{354}{446} }

\ref{Wanders, W. et al. 1995, \ApJ{453}{L87} }
      
\ref{Wanders, W. \& Peterson, B.M. 1996, \ApJ{466}{174} }

\ref{Welsh, W.F. \& Horne, K. 1991, \ApJ{379}{586} }
 
\ref{Weymann, R.J., Scott, J.S., Schiano, A.V.R. \& Christiansen, W.A. 1982,
       \ApJ{262}{497} }

\ref{White, \& Peterson, B.M. 1994, PASP, 106, 879}

\ref{Wilkes, B.J. 1984, MNRAS, 207, 73}
          
\vfill\eject

\line{\bf FIGURE CAPTIONS\hfill}
\bigskip

Fig.~1.  Schematic cross-section of the model BLR in the $rz$-plane. The C~IV 
         emission originates between $3^\circ$ and $30^\circ$ above the disk 
         plane (dotted region). The aspect angle $i$ is the disk inclination to
         the observer. The gray-shaded region, where $\tau_{obs}>1$
         contains dusty molecular part of the wind.

\medskip 
Fig.~2.  The auxiliary function $F(\psi)$ represents the dependence of 
         $\tau_{obs}$ on the polar angle $\psi$ through $\sigma_c$ and $n_c$
         (eq.~3). It is obtained by solving equations of the MHD flow (EBS).
        
\medskip 
Fig.~3a. Plot of the 1460\AA\ continuum window flux used to drive the model
         for the period of the {\it HST} portion of the K95 campaign.

\medskip
Fig.~3b. Plot of the 1350\AA\ continuum window flux used to drive the model
         in the C91 campaign.

\medskip
Fig.~4.  A sample of C~IV line profiles (thick lines) generated by
         the wind model superimposed on the {\it HST} data. The Julian
         dates are shown in the upper right corners. The vertical axis
         is given in relative intensity units. The horizontal axis gives the 
         Doppler shift in velocity units of ${\rm km\ s^{-1}}$ with respect to 
         the line center (negative values mean a blueshift). The jagged line
         represents the observed line profiles.

\medskip
Fig.~5.  Plot of the total line flux and the six line segment fluxes 
         of the data (error bars) and the model (solid lines) as a function of time 
         (JD~2,400,000$+$) during the {\it HST} observing period of the K95 campaign. 
         The velocity ranges
         for each segment are defined in the Table~1b. The thick error bars 
         and curves corresponds to blue side of the C~IV line and the thin error 
         bars and curves corresponds to red.  The fluxes are normalized to units 
         of $10^{-14}\ {\rm ergs\ cm^{-2}\ s^{-1}}$.

\medskip
Fig.~6.  Plot of the total C~IV line flux in the model (line) and the data
         (error bars) as a function of time during the C91 campaign. The flux
	 is normalized to units of $10^{-14}\ {\rm ergs\ cm^{-2}\ s^{-1}}$.
         The absence of the synthetic C~IV light curve before JD~2,407,535
         is due to the finite time interval (25 days) it takes for the continuum
         to propagate through the model (continuum data is available only after
         JD~2,407,510).

\medskip

Fig.~7a. The CCF for the MRW with the K95 continuum. The model CCF is shown
	 as a solid curve and the data CCF is the dashed curve.

\medskip
Fig.~7b. The CCF for the MBW with the MRW. The model CCF is shown
	 as a solid curve and the data CCF is the dashed curve.

\medskip
Fig.~8a. The 2DTF, $\Psi(v,t)$, for the model driven by a rectangular 0.1-day pulse 
         and in the presence of local electron scattering. The contour levels differ 
         by 5\%.

\medskip
Fig.~8b. The 2DTF, $\Psi(v,t)$, for the model driven by a rectangular 0.1-day pulse,
         without electron scattering. The contour levels differ by 5\%.
\medskip
Fig.~9.  Two-dimensional cross-sections of the BLR at different increasing altitudes 
         above the disk. $(a)$ through $(h)$ are at 0.2, 0.4, 0.6, 0.8, 1.0, 2.0,
         4.0 and 8.0 light days above the disk plane. The azimuthal angle in the
         plane of the disk is limited to $90^o-270^o$ because
         only this part of the BLR is directly observable due to the cloud emission 
         anisotropy. The area between $90^o-180^o$ is receeding and the are between
         $180^o-270^o$ is approaching with respect to the observer. The black areas 
         are regions which also cannot be seen by the observer because of the cloud 
         emission anisotropy. The dark-to-light gray scale represents the FRW-to-FBW
         transition (neglecting electron scattering broadening of the line!). The 
         horizontal solid lines represent the inner and outer cylindrical boundaries 
         of the C~IV-emitting BLR at each altitude. The cross-hatched areas of
         non-emitting material are left for convenience only. Above the upper 
         horizontal line $\tau_{obs} >1.0$ while below the lower horizontal line the 
         clouds are over-ionized. Note, that the MRW turns into an `island' and 
         disappears by 2 light days above the disk (see section 5.2 for additional
         explanations).

\medskip
Fig.~10. The 1DTF for the model, obtained by collapsing the 2DTF (Fig.~8a) along 
         the  velocity axis and normalized to the peak at 4 days. The continuous
         line represents the 1DTF using a 0.1-day pulse; the dashed line represents
         the 1DTF convolved with a 3.8-day Gaussian.

\medskip
Fig.~11. Comparison of the mean line profiles of the data and the model during the
         K95 campaign (includes the NLR contribution). The BLR contribution was 
         obtained from the 2DTF (Fig.~8a) collapsed along the time axis.

\medskip

\vfill\eject
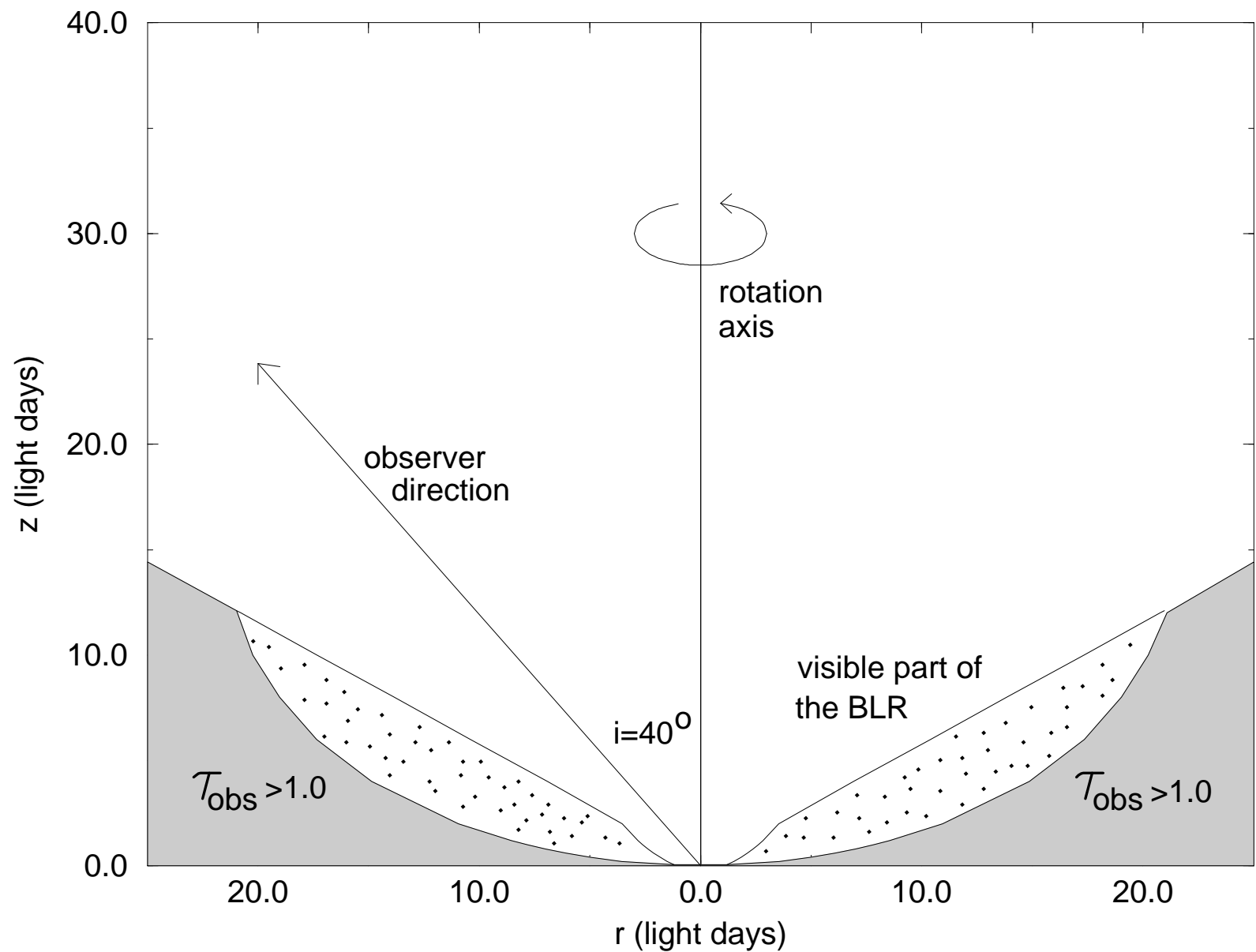
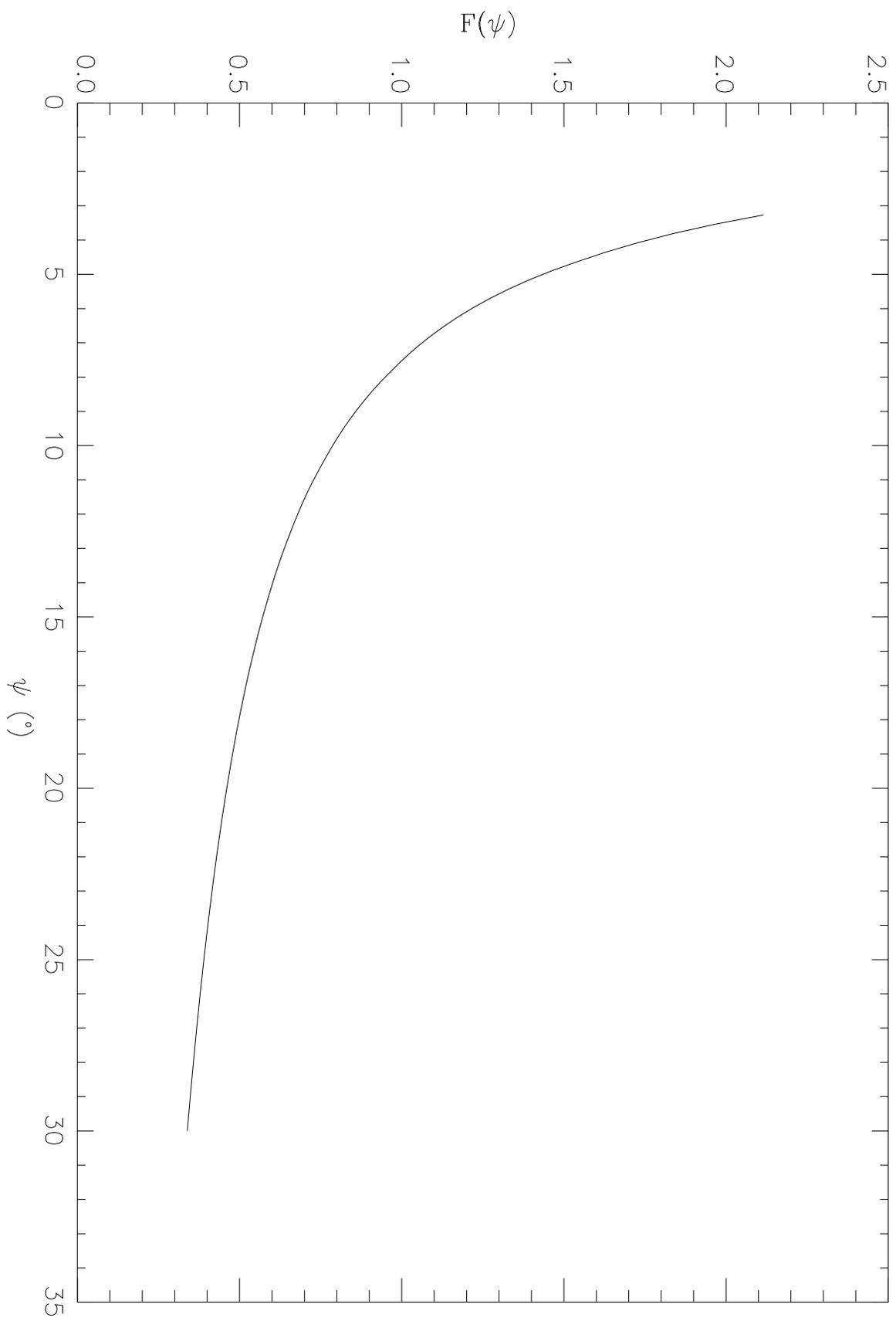
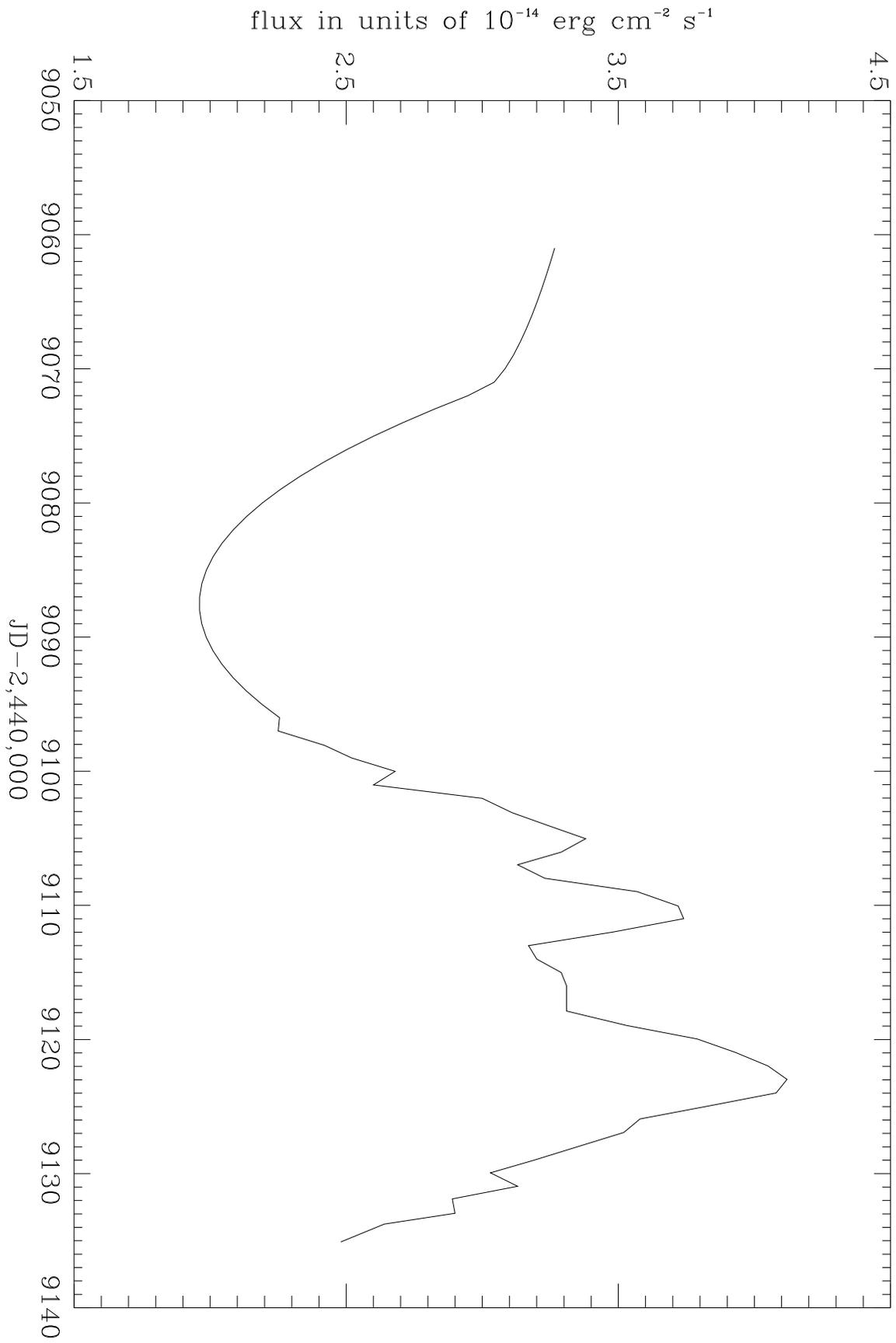
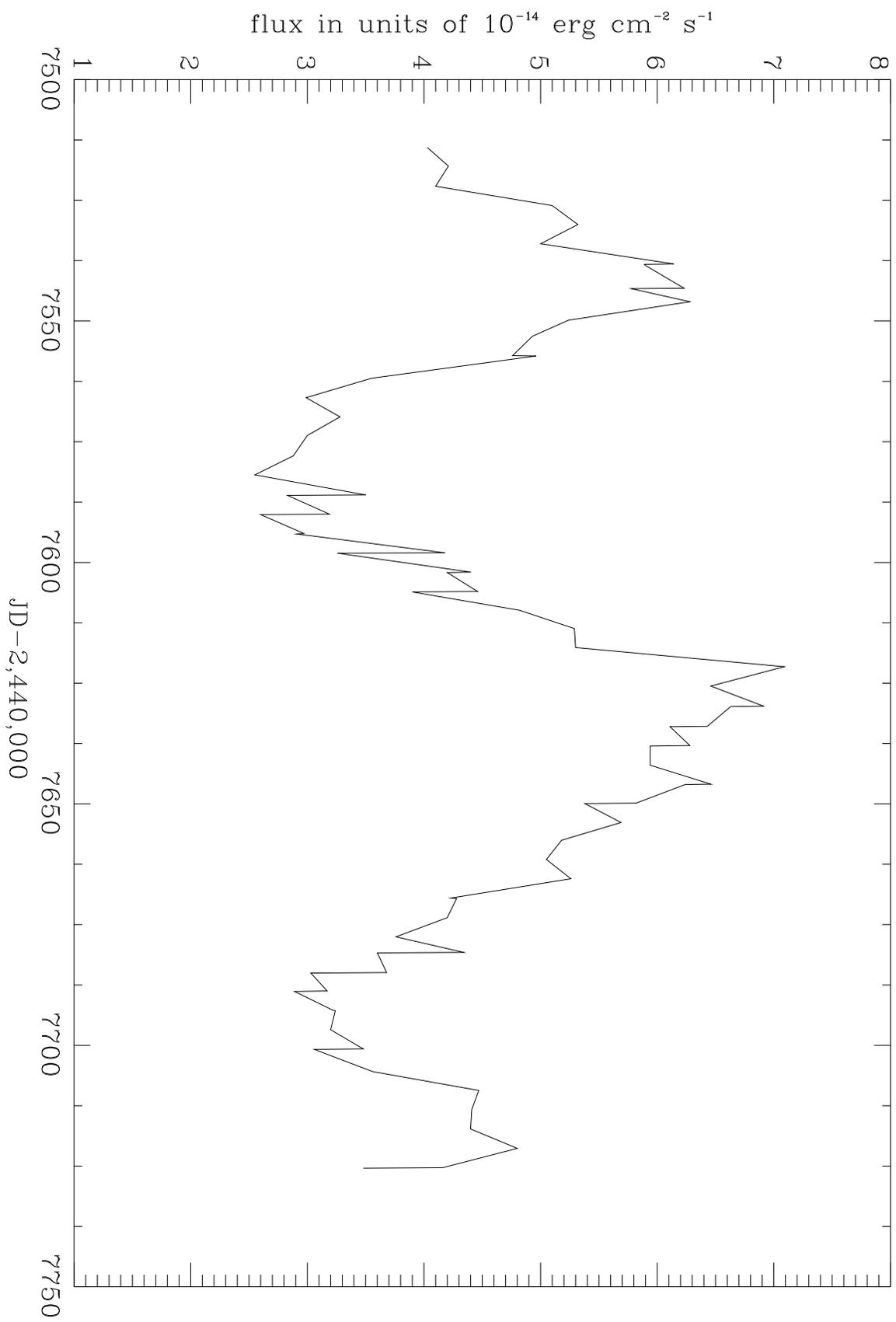
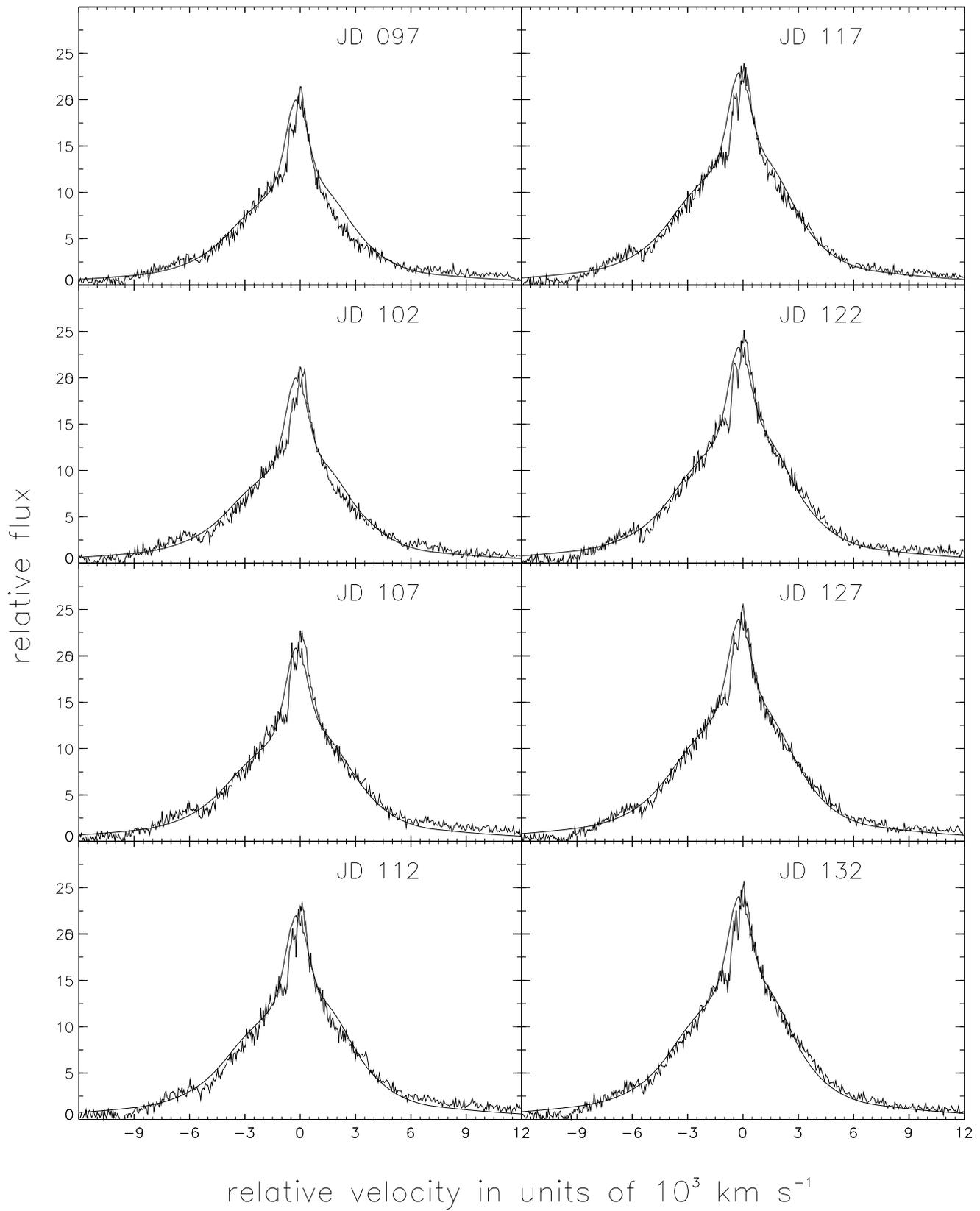
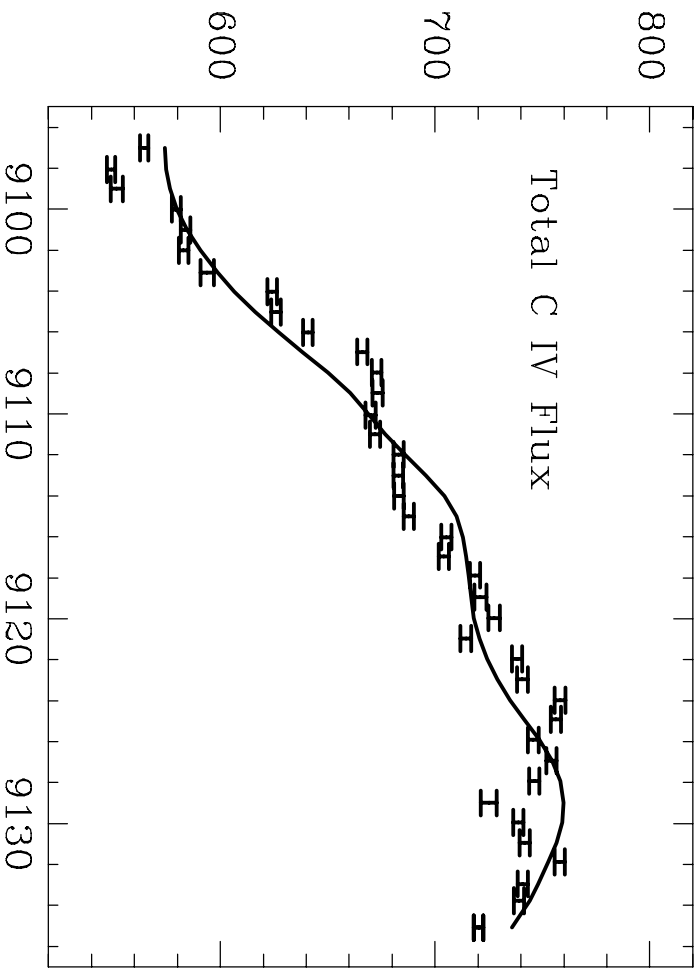
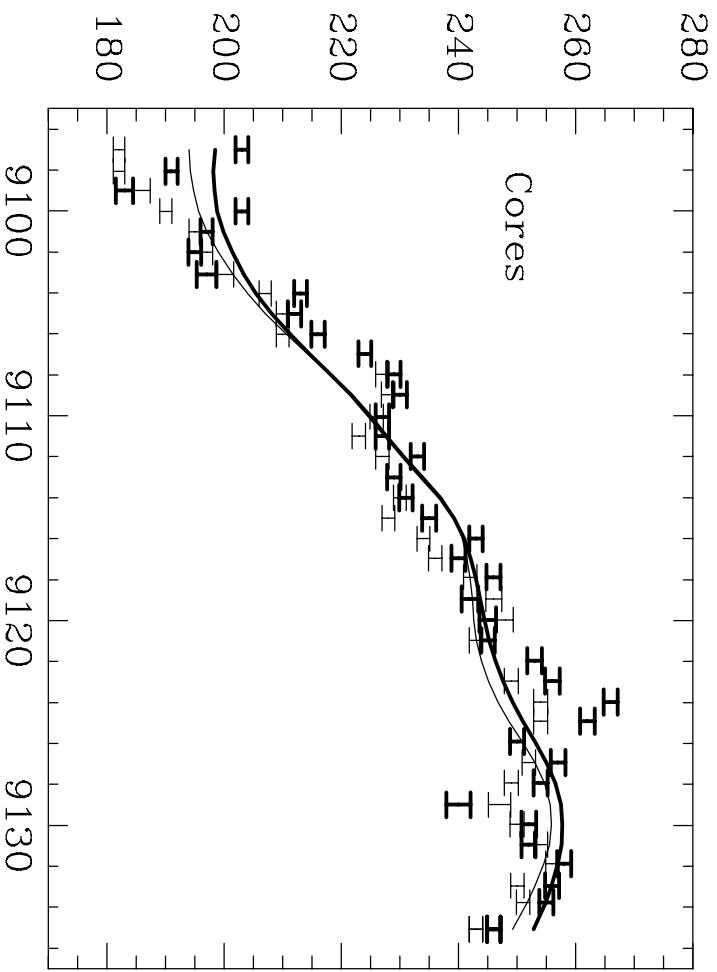
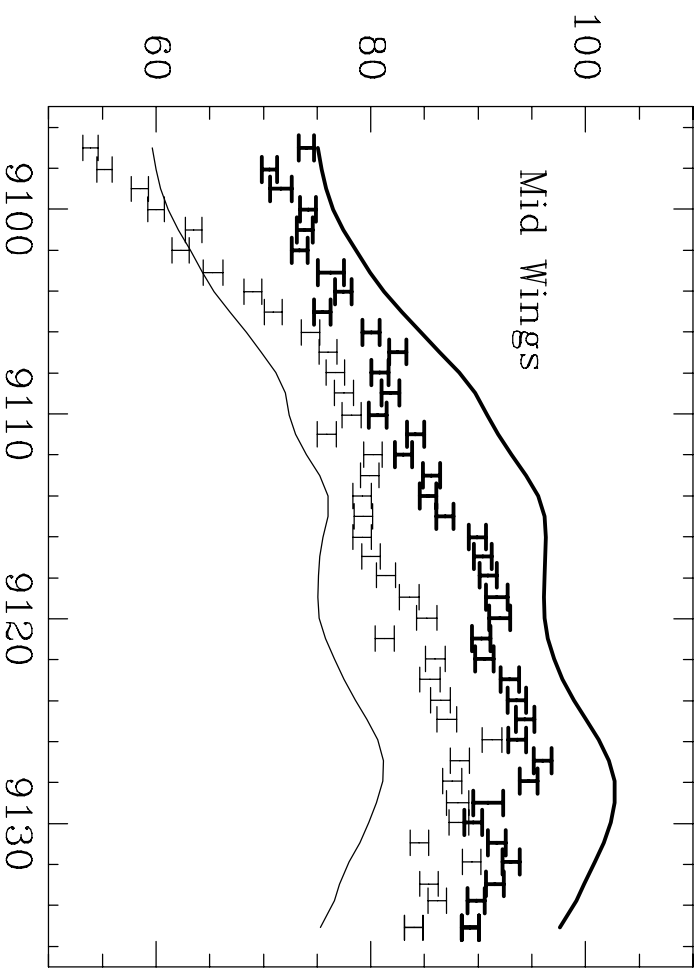
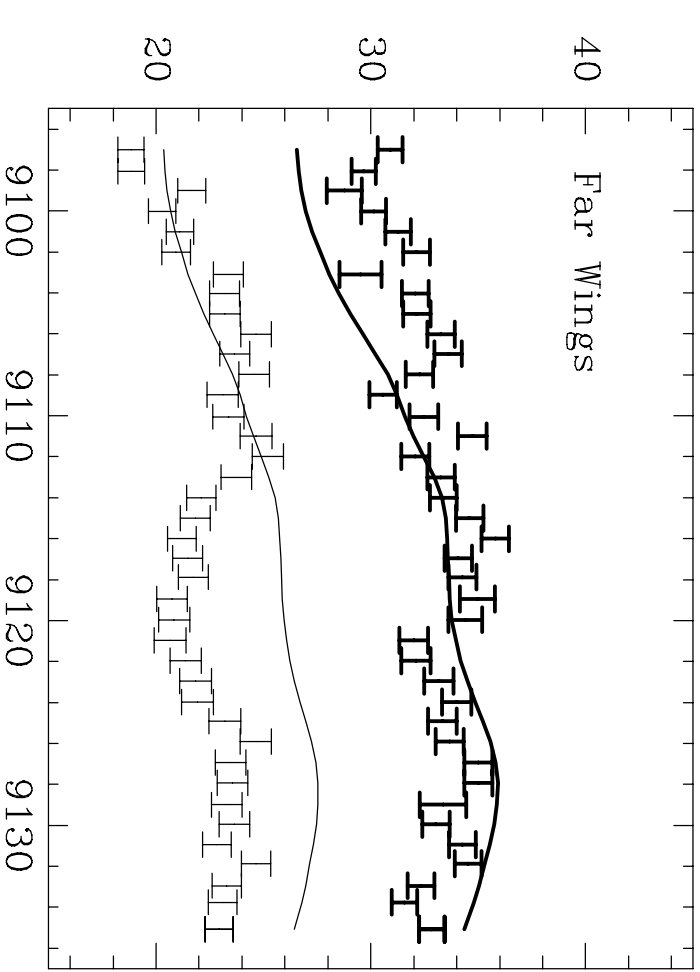
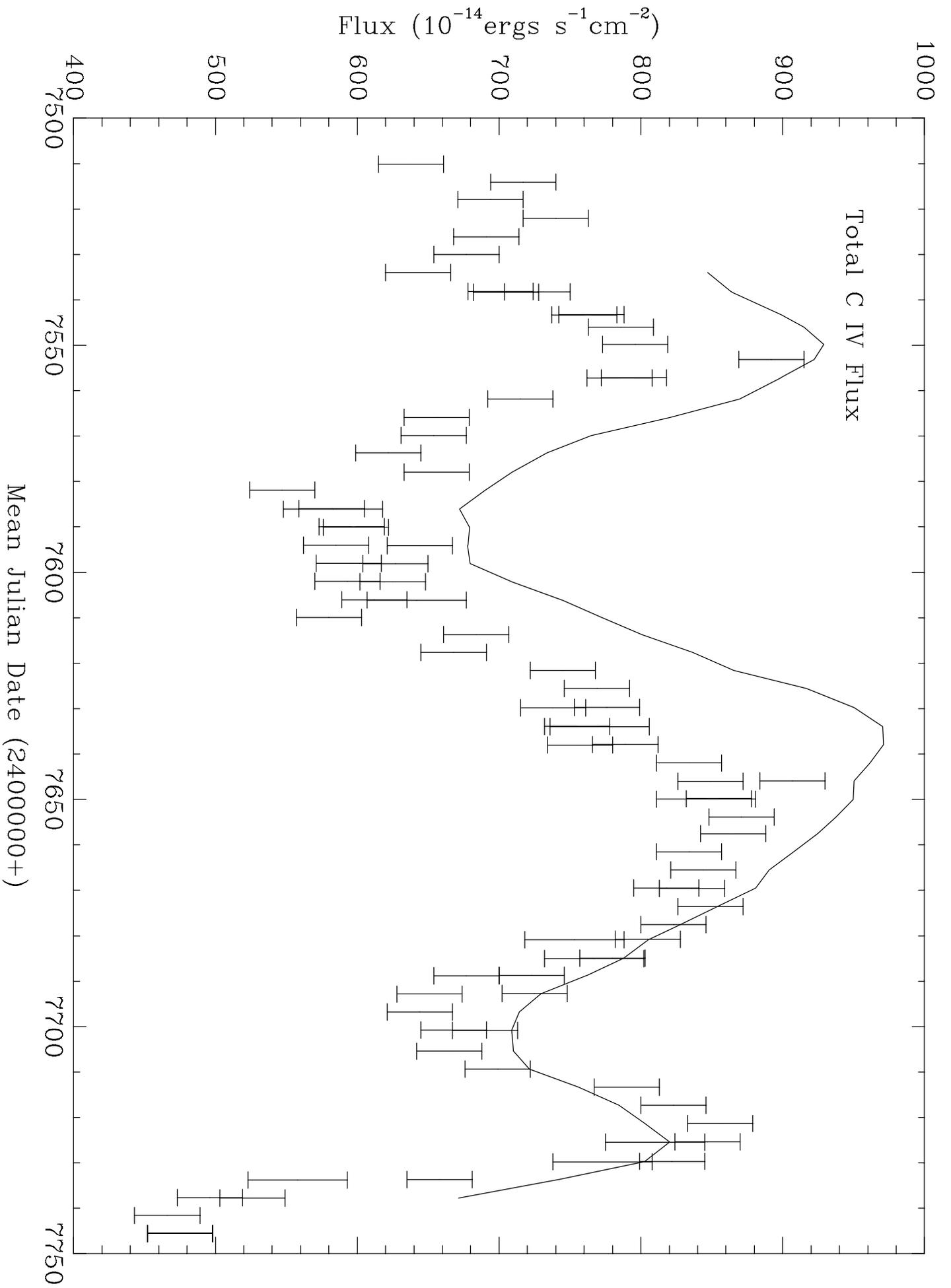
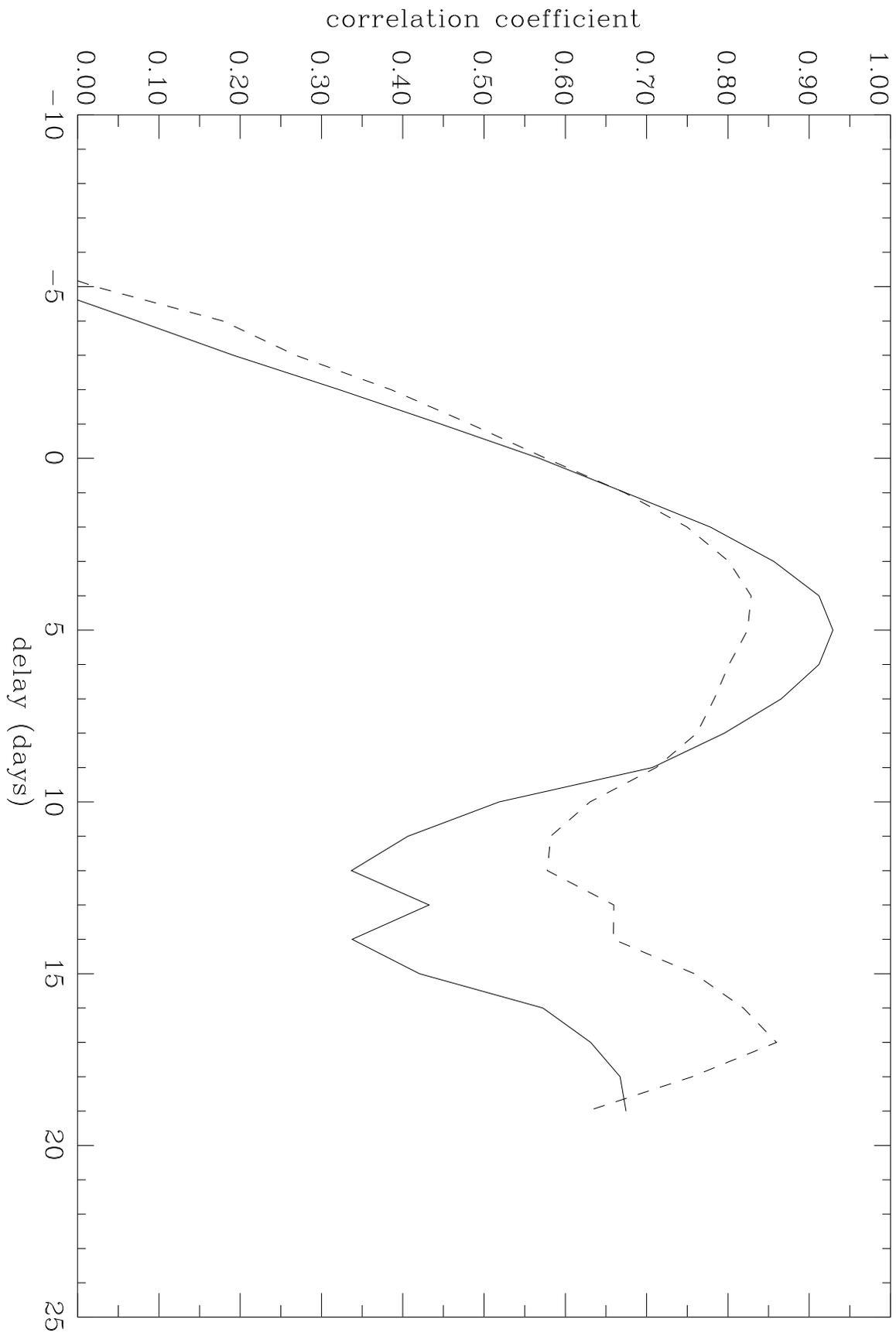
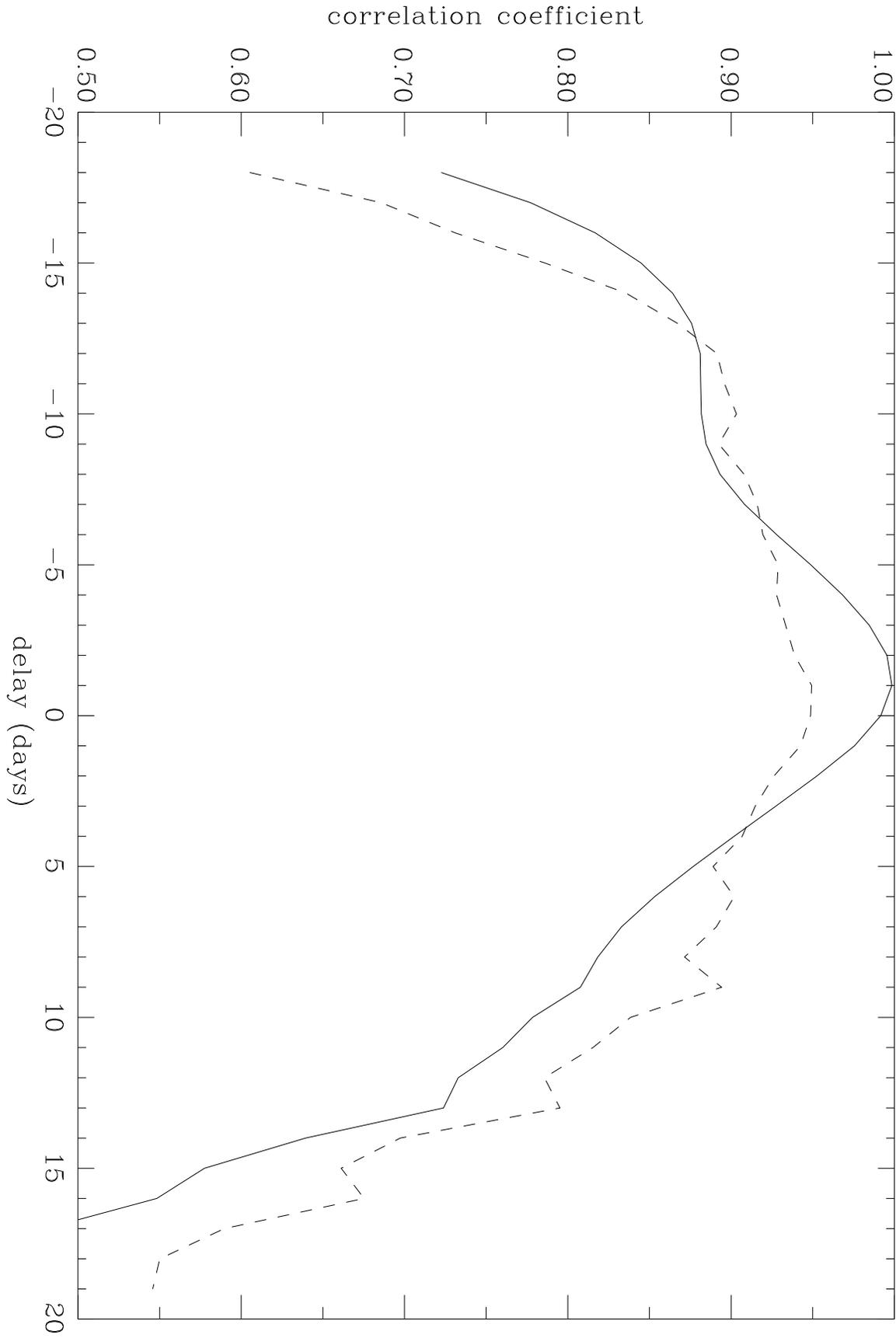
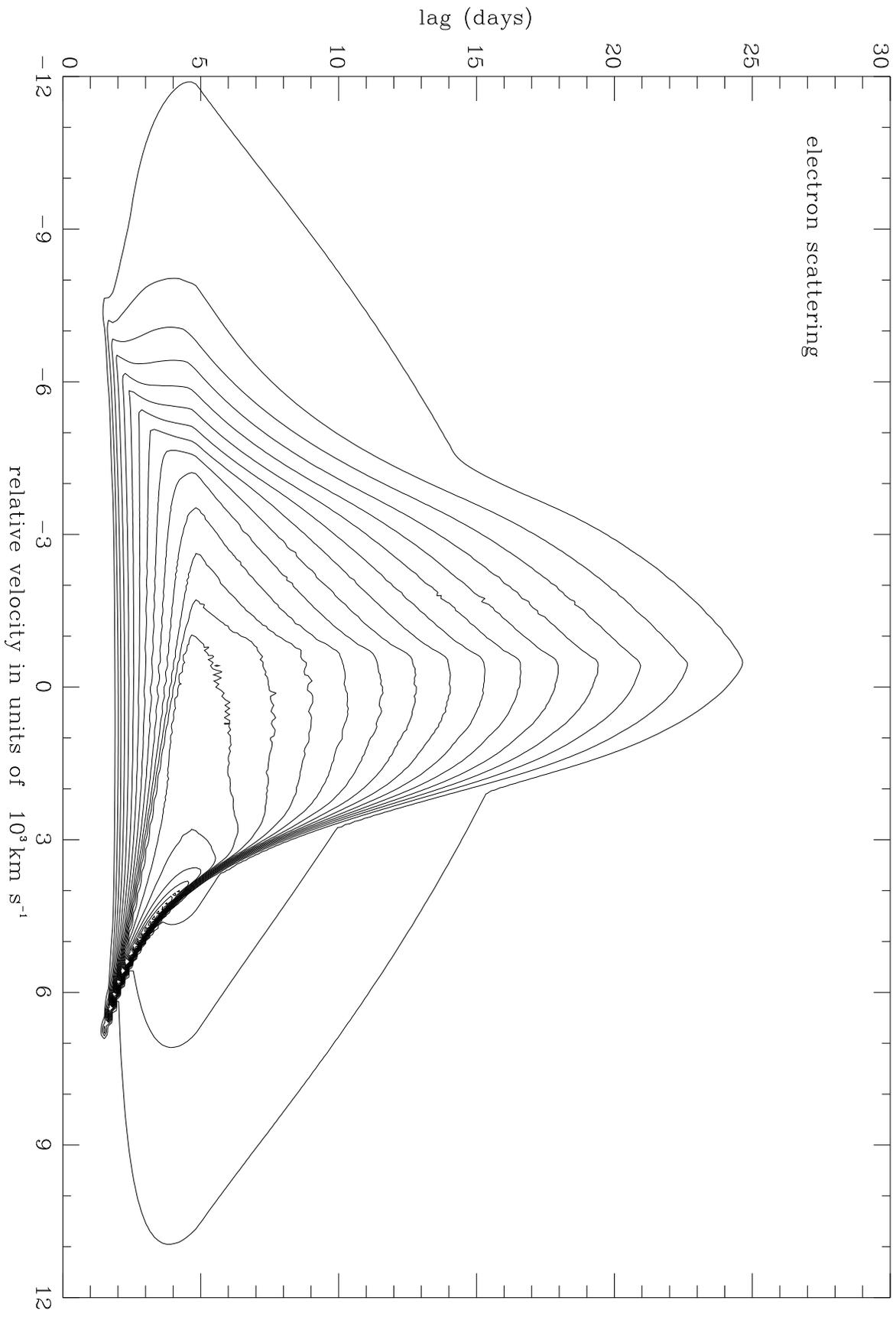
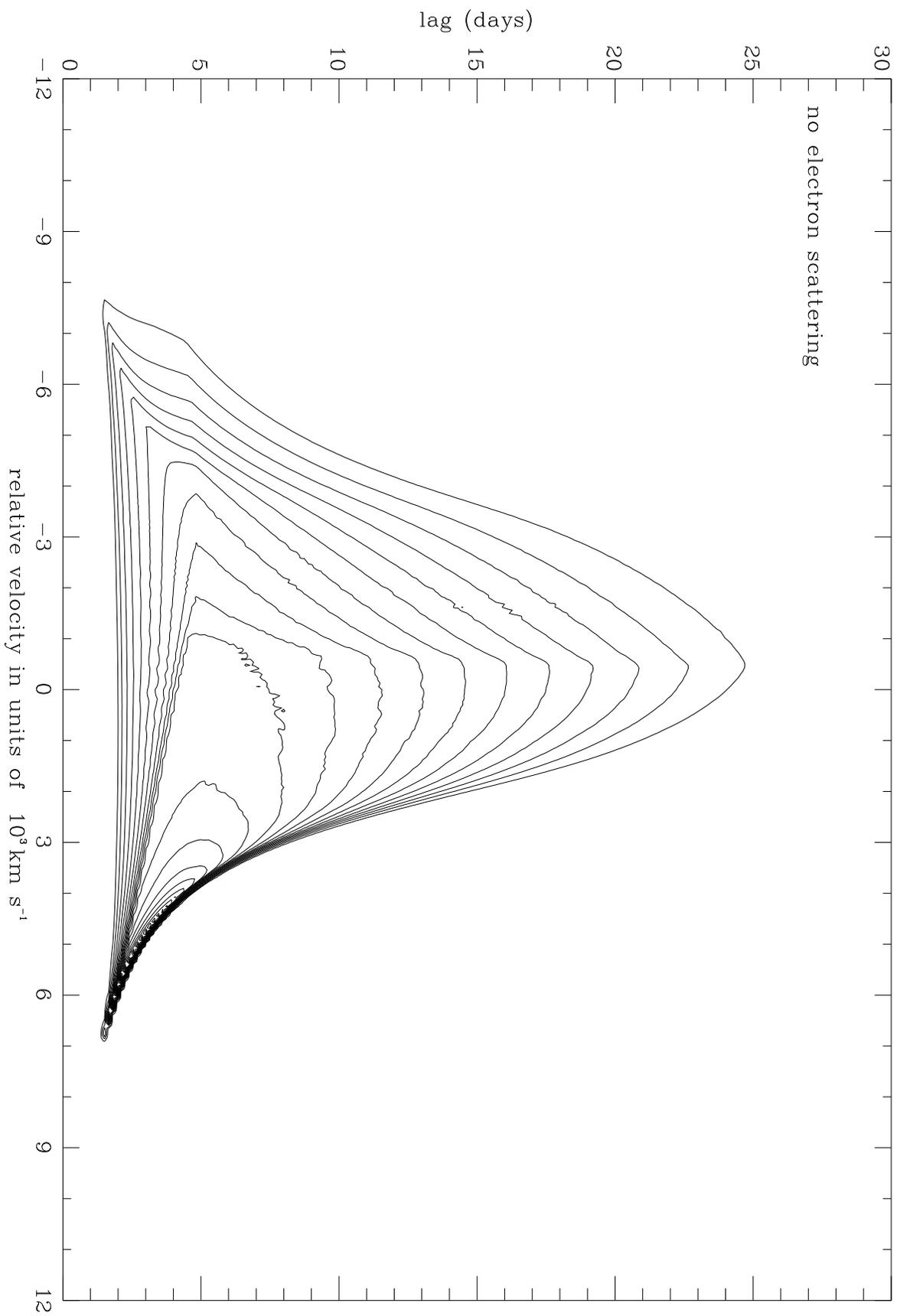
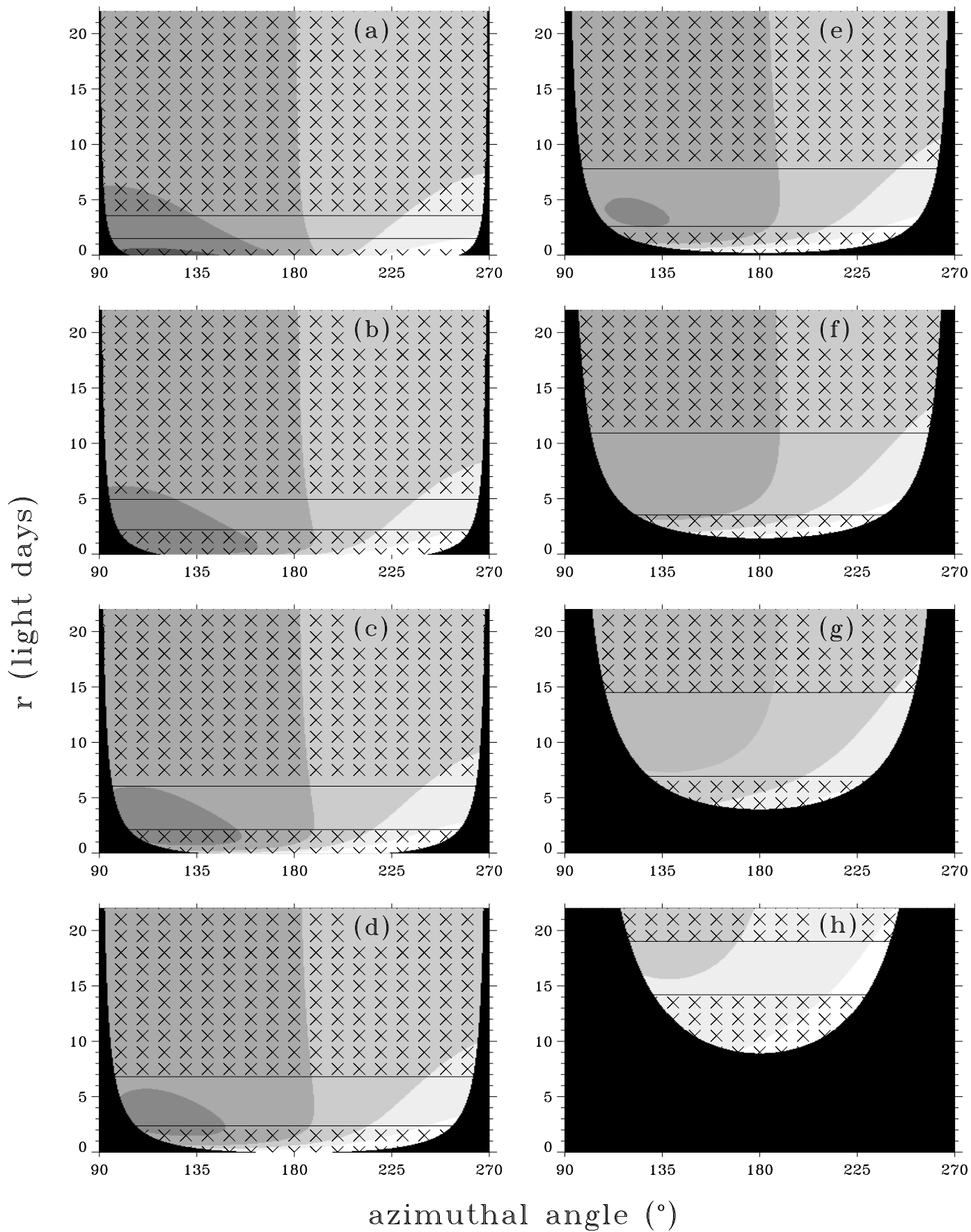
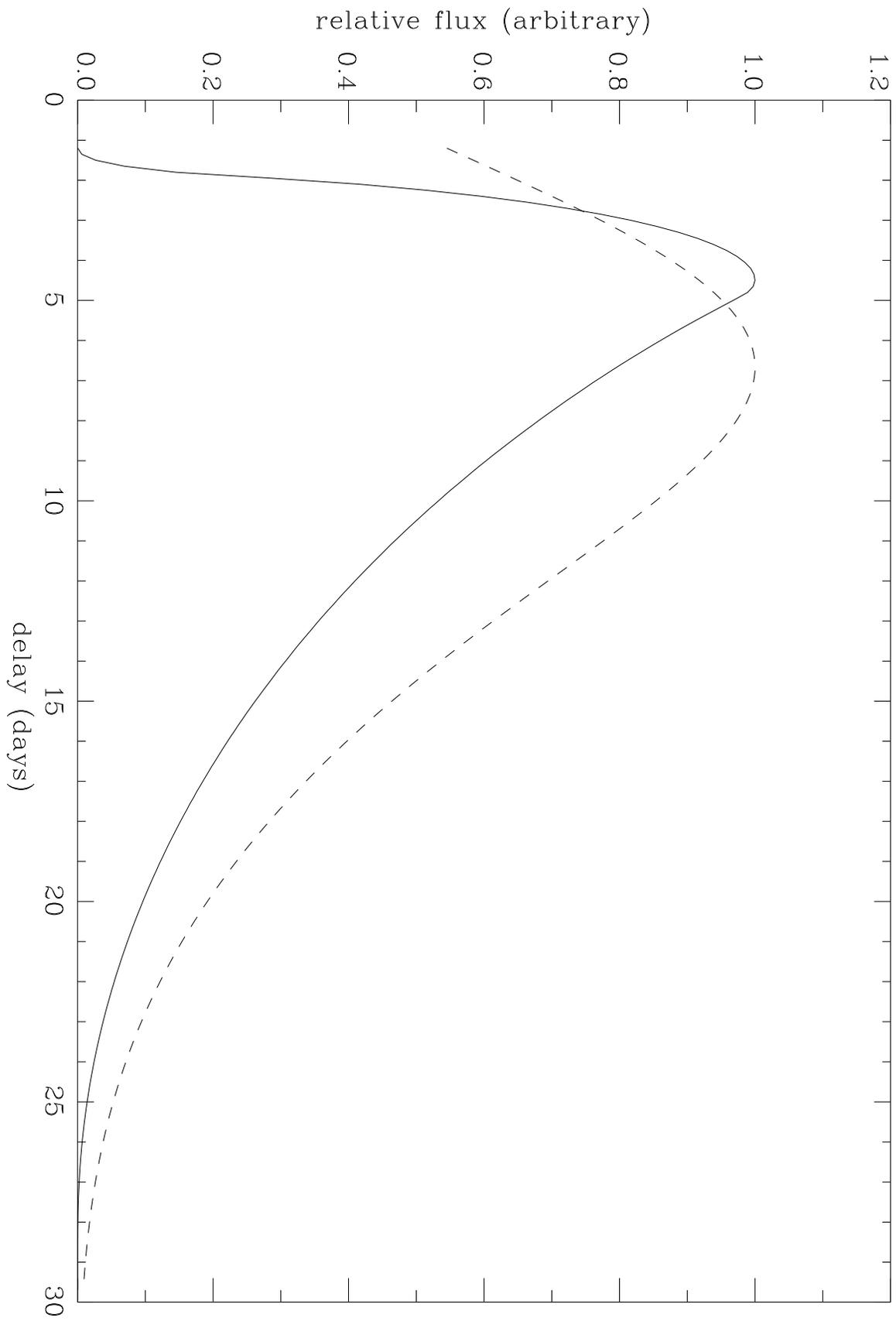
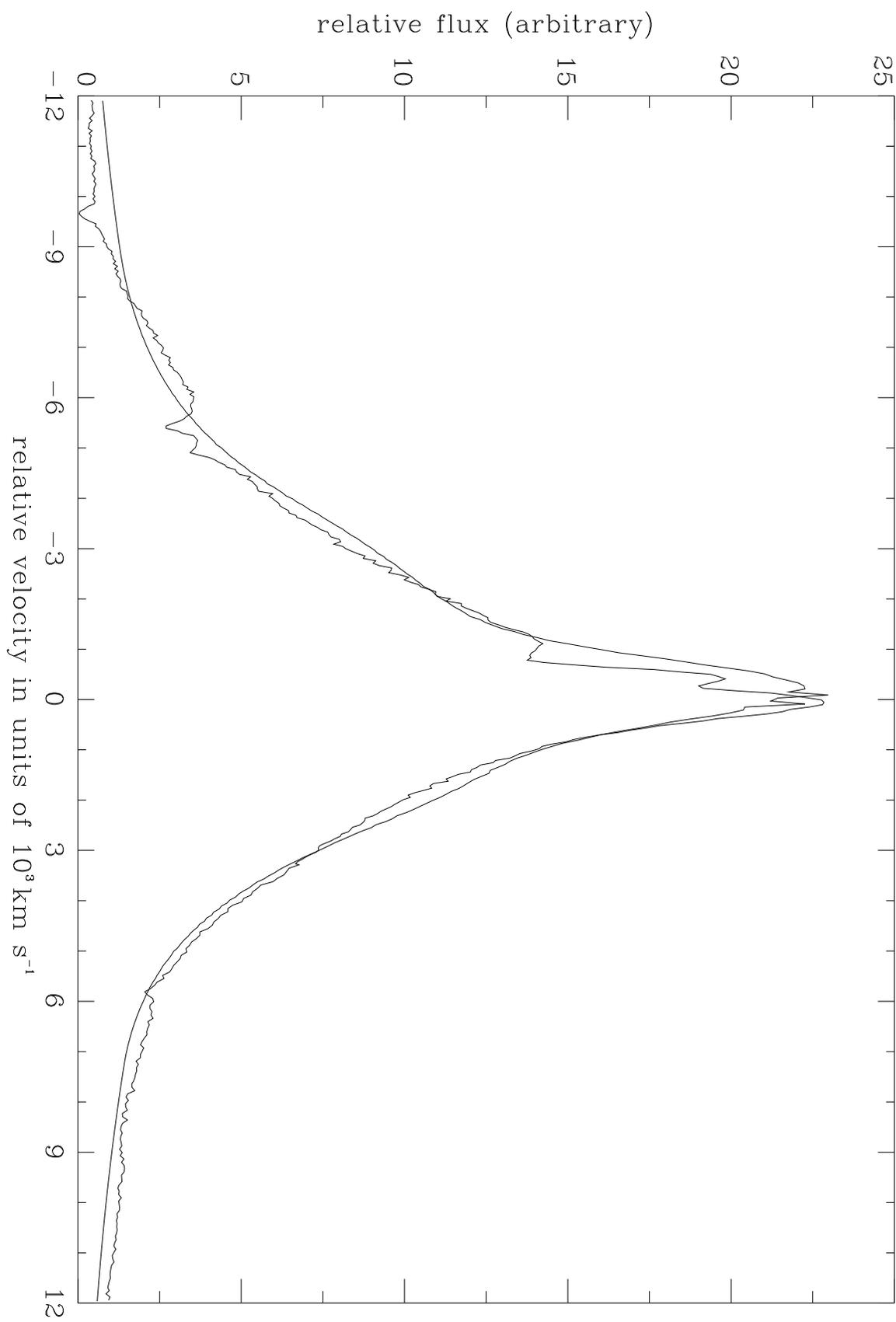
\end